%% file: HIG-17-017_temp.tex
\pdfoutput=1

\documentclass[11pt,twoside,a4paper,cmspaper,final,collab]{cms-tdr}
\def\svnVersion{494004P}\def\cmsCernNoTag{CERN-EP-2018-269}\def\cmsCernDate{\today}\def\cmsMessage{Published in the Journal of High Energy Physics as \href{http://dx.doi.org/10.1007/JHEP04(2019)112}{\doi{10.1007/JHEP04(2019)112.}}}

\begin{document}\cmsNoteHeader{HIG-17-017}

\hyphenation{had-ron-i-za-tion}
\hyphenation{cal-or-i-me-ter}
\hyphenation{de-vices}
\RCS$HeadURL: svn+ssh://svn.cern.ch/reps/tdr2/papers/HIG-17-017/trunk/HIG-17-017.tex $
\RCS$Id: HIG-17-017.tex 494002 2019-04-18 17:58:11Z atiko $
\newlength\cmsFigWidth
\ifthenelse{\boolean{cms@external}}{\setlength\cmsFigWidth{0.85\columnwidth}}{\setlength\cmsFigWidth{0.4\textwidth}}
\ifthenelse{\boolean{cms@external}}{\providecommand{\cmsLeft}{top\xspace}}{\providecommand{\cmsLeft}{left\xspace}}
\ifthenelse{\boolean{cms@external}}{\providecommand{\cmsRight}{bottom\xspace}}{\providecommand{\cmsRight}{right\xspace}}

\ifthenelse{\boolean{cms@external}}
{\providecommand{\suppMaterial}{the supplemental material
[URL will be inserted by publisher]}}
{\providecommand{\suppMaterial}{Appendix~\ref{app:suppMat}}}

\newlength\cmsTabSkip\setlength{\cmsTabSkip}{1ex}
\providecommand{\cmsTable}[1]{\resizebox{\textwidth}{!}{#1}}

\ifthenelse{\boolean{cms@external}}{\providecommand{\CL}{C.L.\xspace}}{\providecommand{\CL}{CL\xspace}}

\newcommand{\mHiggs}{\ensuremath{m_{\PH}}\xspace}
\newcommand{\HH}{\ensuremath{\PH\PH}\xspace}
\newcommand{\mt}{\ensuremath{m_{\PQt}}\xspace}
\newcommand{\fb}{\unit{fb}}
\newcommand{\pp}{\Pp\Pp\xspace}

\newcommand{\HHTobbbb}{\ensuremath{\PH\PH\to\bbbar\bbbar}\xspace}
\newcommand{\ppToHHbbbb}{\ensuremath{\pp\to\HHTobbbb}\xspace}
\newcommand{\HTobb}{\ensuremath{\PH\to\bbbar}\xspace}

\newcommand{\xsppToHHSM}{\ensuremath{\sigma{(\pp\to\HH)_\text{SM}}}\xspace}
\newcommand{\xsppToHHbbbb}{\ensuremath{\sigma{(\ppToHHbbbb)}}\xspace}

\newcommand{\bbbb}{\ensuremath{\bbbar\bbbar}\xspace}
\newcommand{\tttt}{\ensuremath{\ttbar\ttbar}\xspace}
\newcommand{\bblnulnu}{\ensuremath{\bbbar\ell\PGn\ell\PGn}\xspace}
\newcommand{\bblnuqq}{\ensuremath{\bbbar\ell\PGn\Pq\Pq}\xspace}
\newcommand{\bbtautau}{\ensuremath{\bbbar\PGt\PGt}\xspace}
\newcommand{\bbgammagamma}{\ensuremath{\bbbar\PGg\PGg}\xspace}
\newcommand{\gammagammaWW}{\ensuremath{\PGg\PGg\PW\PW^{*}}\xspace}
\newcommand{\WWWW}{\ensuremath{\PW\PW^{*}\PW\PW^{*}}\xspace}
\newcommand{\bbH}{\ensuremath{\bbbar\PH}\xspace}
\newcommand{\ttH}{\ensuremath{\ttbar\PH}\xspace}
\newcommand{\ttbb}{\ensuremath{\ttbar\bbbar}\xspace}
\newcommand{\ZH}{\ensuremath{\PZ\PH}\xspace}

\newcommand{\ggF}{\ensuremath{\cPg\cPg\text{F}}\xspace}

\newcommand{\kappalambda}{\ensuremath{\kappa_{\text{\Lam}}}\xspace}
\newcommand{\kappat}{\ensuremath{\kappa_{\cPqt}}\xspace}
\newcommand{\kappab}{\ensuremath{\kappa_{\cPqb}}\xspace}
\newcommand{\ctwo}{\ensuremath{c_\mathrm{2}}\xspace}
\newcommand{\cg}{\ensuremath{c_{\cPg}}\xspace}
\newcommand{\ctwog}{\ensuremath{c_{2\cPg}}\xspace}
\newcommand{\lambdaSM}{\ensuremath{\Lam_{\text{SM}}}\xspace}
\newcommand{\lambdaHHH}{\ensuremath{\Lam_{\PH\PH\PH}}\xspace}
\newcommand{\ySMt}{\ensuremath{y_{\text{SM}}}\xspace}
\newcommand{\yt}{\ensuremath{y_{\cPqt}}\xspace}

\newcommand{\tltr}{\ensuremath{\overline{\cmsSymbolFace{t}}_\cmsSymbolFace{L} \cmsSymbolFace{t}_\cmsSymbolFace{R}}}

\newcommand{\hc}{\ensuremath{\text{h.c.}}}

\newcommand{\PHone}{\ensuremath{\PH_1}\xspace}
\newcommand{\PHtwo}{\ensuremath{\PH_2}\xspace}

\newcommand{\MHH}{\ensuremath{M_{\PH\PH}}\xspace}
\newcommand{\MX}{\ensuremath{M_{\mathrm{X}}}\xspace}
\newcommand{\MHone}{\ensuremath{M_{\PHone}}\xspace}
\newcommand{\MHtwo}{\ensuremath{M_{\PHtwo}}\xspace}
\newcommand{\Mfourj}{\ensuremath{m_{4\mathrm{j}}}\xspace}

\newcommand{\pTHH}{\ensuremath{\pt^{\PHone\PHtwo}}\xspace}
\newcommand{\pTHone}{{\ensuremath{\pt^{\PHone}}}\xspace}
\newcommand{\pTHtwo}{{\ensuremath{\pt^{\PHtwo}}}\xspace}

\newcommand{\mHHgen}{\ensuremath{m_{\HH}^{\text{gen}}}\xspace}

\newcommand{\thetastar}{\ensuremath{\theta^{*}}\xspace}
\newcommand{\ctstarGen}{\ensuremath{\cos\thetastar_{\text{gen}}}\xspace}

\newcommand{\ptJet}{\ensuremath{{\pt}_{\mathrm{j}}}\xspace}
\newcommand{\absEtaJet}{\ensuremath{\abs{\eta_{\mathrm{j}}}}\xspace}

\newcommand{\pTjets}{\ensuremath{\ptJet^{i}}\xspace}
\newcommand{\etaJets}{\ensuremath{\eta^{i}_{\mathrm{j}}}\xspace}
\newcommand{\pTjetsFour}{\ensuremath{\ptJet^{(i = 1\text{--}4)}}\xspace}
\newcommand{\etaJetsFour}{\ensuremath{\eta^{(i = 1\text{--}4)}_{\text{j}}}\xspace}
\newcommand{\pTjetind}[1]{\ensuremath{\ptJet^{#1}}\xspace}
\newcommand{\etaJetind}[1]{\ensuremath{\eta^{#1}_{\text{j}}}\xspace}

\newcommand{\Nj}{\ensuremath{N_\text{j}}\xspace}
\newcommand{\Nb}{\ensuremath{N_\text{b}}\xspace}
\newcommand{\Mtot}{\ensuremath{M_\text{tot}}\xspace}

\newcommand{\Njind}[1]{\ensuremath{N_\text{j}^{#1}}\xspace}
\newcommand{\Nbind}[1]{\ensuremath{N_\text{b}^{#1}}\xspace}
\newcommand{\Mtotind}[1]{\ensuremath{M_\text{tot}^{#1}}\xspace}
\newcommand{\Thrust}{\ensuremath{T}\xspace}
\newcommand{\Tind}[1]{\ensuremath{T^{#1}}\xspace}
\newcommand{\Taind}[1]{\ensuremath{T_{a}^{#1}}\xspace}
\newcommand{\sumpzind}[1]{\ensuremath{\Sigma p_{z}^{#1}}\xspace}

\newcommand{\Dist}[2]{{\ensuremath{D(#1,#2)}}\xspace}
\newcommand{\Ta}{\ensuremath{T_{a}}\xspace}
\newcommand{\sumpz}{\ensuremath{\Sigma p_{z}}\xspace}

\newcommand{\Njh}{\Njind{h}}
\newcommand{\Nbh}{\Nbind{h}}
\newcommand{\Mtoth}{\Mtotind{h}}

\newcommand{\Th}{\Tind{h}}
\newcommand{\Tah}{\Taind{h}}
\newcommand{\sumpzh}{\sumpzind{h}}

\newcommand{\deltaMklmn}{\ensuremath{\Delta \cmsSymbolFace{M}_{klmn}}\xspace}
\newcommand{\diffMklmn}{\ensuremath{\abs{\cmsSymbolFace{M}_{kl} - \cmsSymbolFace{M}_{mn}}}\xspace}

\newcommand{\Deta}{\ensuremath{\Delta\eta}\xspace}
\newcommand{\DPhi}{\ensuremath{\Delta\phi}\xspace}

\newcommand{\DRjjHone}{\ensuremath{\Delta R_\mathrm{jj}^{\PHone}}\xspace}
\newcommand{\DRjjHtwo}{\ensuremath{\Delta R_\mathrm{jj}^{\PHtwo}}\xspace}
\newcommand{\DPhijjHone}{\ensuremath{\DPhi_\mathrm{jj}^{\PHone}}\xspace}
\newcommand{\DPhijjHtwo}{\ensuremath{\DPhi_\mathrm{jj}^{\PHtwo}}\xspace}

\newcommand{\ctsHoto}{\ensuremath{\cos \thetastar_{\PHone \PHtwo-\PHone}}\xspace}
\newcommand{\ctsHoneJone}{\ensuremath{\cos \thetastar_{\PHone\text{--}\mathrm{j}_1}}\xspace}
\newcommand{\HTrest}{\ensuremath{\HT^\text{rest}}\xspace}

\newcommand{\muF}{\ensuremath{\mu_\mathrm{F}}\xspace}
\newcommand{\muR}{\ensuremath{\mu_\mathrm{R}}\xspace}

\newcommand {\sd}{\unit{s.d.}}
\newcommand{\cmva}[1]{CMVA_{#1}\xspace}

\newcommand{\HTgen}{\ensuremath{\HT^\text{gen}}\xspace}

\cmsNoteHeader{HIG-17-017}

\date{\today}
\title{Search for nonresonant Higgs boson pair production in the \bbbb final state at $\sqrt{s} = 13\TeV$}

\abstract{
Results of a search for nonresonant production of Higgs boson pairs, with each Higgs boson decaying to a \bbbar pair, are presented. This search uses data from proton-proton collisions at a centre-of-mass energy of 13\TeV, corresponding to an integrated luminosity of $35.9\fbinv$, collected by the CMS detector at the LHC. No signal is observed, and a 95\% confidence level upper limit of 847\fb is set on the cross section for standard model nonresonant Higgs boson pair production times the squared branching fraction of the Higgs boson decay to a \bbbar pair. The same signature is studied, and upper limits are set, in the context of models of physics beyond the standard model that predict modified couplings of the Higgs boson.
}
\hypersetup{
pdfauthor={CMS Collaboration},
pdftitle={Search for nonresonant Higgs boson pair production in the bbbb final state at sqrt(s) = 13 TeV},
pdfsubject={CMS},
pdfkeywords={CMS, physics, Higgs}}

\maketitle

\section{Introduction}
The detailed understanding of the properties of the Higgs boson (\PH) discovered in 2012
by the CERN LHC experiments~\cite{Aad20121,Chatrchyan201230,Chatrchyan:2013lba}
remains an important subject in fundamental physics.
Current determinations of the properties of the new particle by the ATLAS and CMS Collaborations are found
to be in agreement with standard model (SM) predictions~\cite{higgs_comb, higgs_mass_comb}.
However, there are still many measurements that could reveal unexpected deviations from the SM.
A number of models of physics beyond the SM (BSM) can be tested using their predictions of the properties of the observed state,
including the Higgs boson self-coupling and couplings to bosons and fermions~\cite{Dib:2005re, Grober:2010yv, Contino:2012xk, Dolan:2012ac, Dawson:2015oha}.

The production of Higgs boson
pairs (\HH) is the most direct way to access the Higgs boson self-coupling~\cite{Baglio:2012np}
and to study in detail the SM Higgs potential.
The \HH production cross section predicted by the
SM for 13\TeV proton-proton (\pp) collisions and $\mHiggs = 125.09\GeV$~\cite{higgs_mass_comb,Sirunyan:2017exp}
is $33.49^{+4.3\%}_{-6.0\%} (\text{scale})
\pm 2.3\% (\alpS) \pm 2.1\% (\text{PDF})\fb$~\cite{deFlorian:2016spz,
deFlorian:2013jea,Dawson:1998py,PhysRevLett.117.012001,deFlorian:2015moa},
where the uncertainty is due to the variation of the renormalization (\muR) and factorization (\muF) scales (scale),
the strong coupling constant (\alpS) uncertainties, and
the uncertainty in parton distribution functions (PDF).
The predicted cross section results in a low expected event rate,
and the acceptance for \HH events in the detector is small.
This means that the SM \HH production process cannot be observed
with the data collected so far at the LHC:
the expectation is that it will only be possible to set an upper limit on the \HH production cross section,
as discussed, \eg in Refs.~\cite{2018101bbtautau,Aaboud:2018knk}.
However, the cross section can be enhanced by anomalous couplings in BSM models~\cite{Carvalho:2016rys} and
in some cases the enhancement is large enough that \HH production could be observed with the current data.

The first searches for nonresonant \HH production were performed by LHC experiments using \pp collisions data at
$\sqrt{s} = 8\TeV$~\cite{Aad2015HH4b,Sirunyan:2017tqo}.
The data collected in 2015 and 2016 at $\sqrt{s} = 13\TeV$ were used for improved analyses in the decay channels:
\bbbb~\cite{Aaboud:2018knk},
\bblnulnu~\cite{hh4bbww},
\bblnuqq~\cite{Aaboud:2018zhh},
\bbtautau~\cite{2018101bbtautau, Aaboud:2018sfw}, 
\bbgammagamma~\cite{Sirunyan:2621190, Aaboud:2018ftw},
\gammagammaWW~\cite{Aaboud:2018ewm}, and
\WWWW~\cite{Aaboud:2018ksn}.
An additional search in the \bbbb decay channel focused on the region of phase space where one \bbbar pair is highly Lorentz-boosted and is reconstructed as a single large-area jet~\cite{Sirunyan:2018qca}.
In the cases mentioned above, at least one of the two Higgs bosons is required to decay to \bbbar to exploit the large branching fraction of this decay.
Results were found to be compatible within uncertainties to the expected SM background contribution.
The measurement of nonresonant \HH production at the LHC with the tightest expected upper limit (15 times the SM rate) was made in the \bbtautau
channel~\cite{Aaboud:2018sfw}, yielding an observed upper limit equivalent to 13 times the SM rate.

This article reports the results of a search for \HH production with both Higgs bosons
decaying into bottom quark pairs, resulting in four resolved hadronic jets.
The search is performed using 13\TeV \pp collisions data corresponding to an integrated luminosity of 35.9\fbinv, collected by the CMS detector in 2016.
The final state containing four \cPqb\ quarks has the highest branching fraction of all possible \HH\ final states,
corresponding to ${\approx}0.339$ for an SM Higgs boson with a mass of 125\GeV.
It is one of the most sensitive signatures for the investigation of \HH production,
as confirmed by the results of a similar search recently performed by the ATLAS Collaboration~\cite{Aaboud:2018knk}.
The main challenge for this analysis is the large background from multijet
final states produced by quantum chromodynamics (QCD) processes, which collectively yield rates
exceeding that of the signal by several orders of magnitude.
We address this by fully exploiting the distinctive features of the signal: the presence of four \cPqb\ quarks and the
kinematical properties of the decay process.
In a sample selected by requiring four \cPqb\ quark jets, a multivariate event classifier is trained to discriminate signal from background.
This sample is studied by comparing it to a model of all contributing background processes, which is completely based on data.
Because of the use of different data sets, triggers, and offline selection requirements, this analysis is fully independent from the CMS searches
mentioned above~\cite{hh4bbww, 2018101bbtautau, Sirunyan:2621190,Sirunyan:2018qca}.

\section{Beyond-the-standard-model extensions}
\label{sec:bsm}

\begin{figure}[h]
\centering
\includegraphics[scale=0.85]{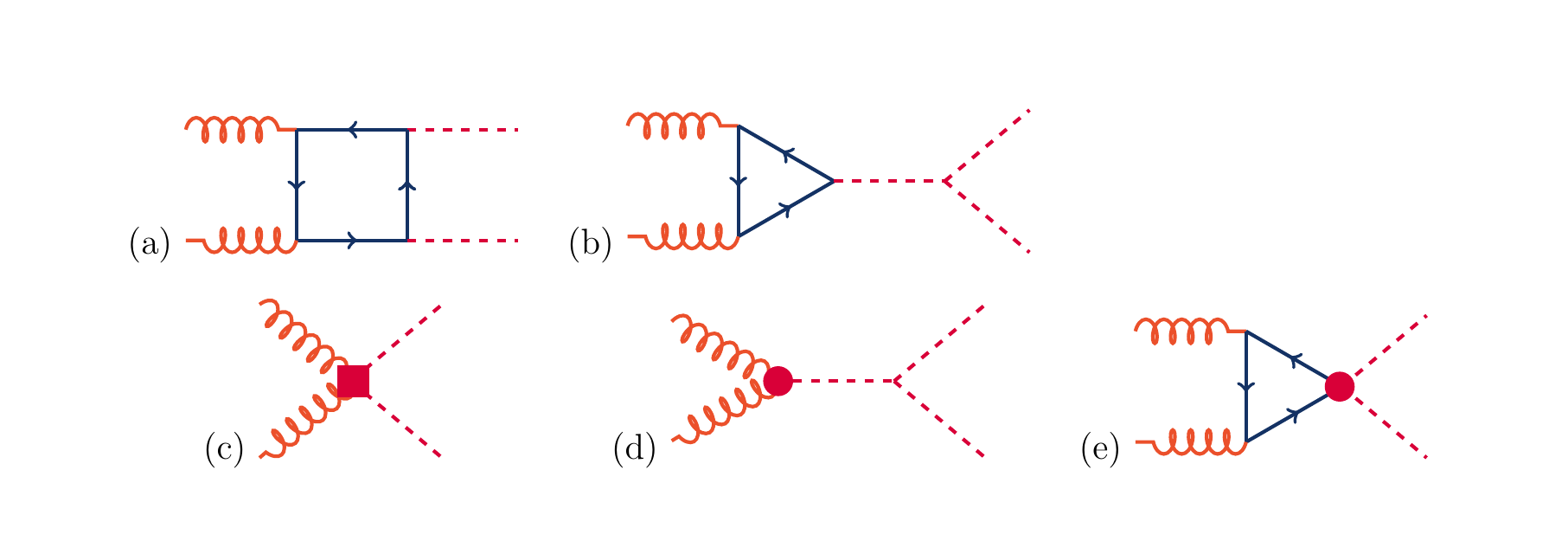}
\caption{Feynman diagrams that contribute to \HH production via gluon-gluon fusion at LO. Diagrams (a) and (b) correspond to SM-like processes,
while diagrams (c), (d), and (e) correspond to pure BSM effects:
(c) and (d) describe contact interactions between the Higgs boson and gluons, and (e) describes the contact interaction of two Higgs bosons with top quarks.   \label{fig:dia}}
\end{figure}

In the SM, \HH production occurs predominantly by gluon-gluon fusion
 (\ggF) via an internal fermion loop, where the top quark (\cPqt) contribution is dominant.
In the absence of new light states, the \ggF\ \HH production at the LHC can be generally described
(considering operators up to dimension 6) by five parameters controlling the tree-level interactions of the Higgs boson.
Without considering CP violating effects, the relevant part of the Lagrangian then takes the form:
\begin{linenomath}
\begin{equation}
\label{eq:lag}
  \begin{aligned}
    {\cal L}_\PH =
\frac{1}{2} \partial_{\mu}\, \PH \, \partial^{\mu} \PH - \frac{1}{2} {\mHiggs}^2 \PH^2 -
  \kappalambda \,  \lambdaSM v\, \PH^3
- \frac{\mt}{v}(v+   {\kappat} \,   \PH  +  \frac{\ctwo}{v}   \HH ) \,( \tltr + \hc) \\
+ \frac{1}{4} \frac{\alpS}{3 \pi v} (   \cg \, \PH -  \frac{\ctwog}{2 v} \,  \HH ) \,  G^{\mu \nu}G_{\mu\nu}\,.
  \end{aligned}
\end{equation}
\end{linenomath}
This Lagrangian follows from extending the SM with operators of mass dimension between four and six in the framework of an effective field theory~\cite{FALKOWSKI2016},
encoding the effects of new heavy states currently beyond experimental reach.
The five parameters of the Lagrangian, named \kappalambda, \kappat, \cg, \ctwog, and \ctwo, are related to the Higgs boson couplings.
In particular, the multiplicative factors
$\kappalambda = \lambdaHHH/\lambdaSM$ and $\kappat = \yt/\ySMt$
parametrize deviations from the SM values of, respectively,
the Higgs boson trilinear coupling and the top quark Yukawa coupling.
The former is given by $\lambdaSM = \mHiggs^2/2v^2$, with $v$ being the vacuum expectation value of the Higgs field.
The absolute couplings $\cg$, $\ctwog$, and $\ctwo$ parametrize contact interactions not predicted by the SM,
\ie the coupling of the Higgs boson to gluons and those of the  two Higgs bosons to two gluons or to a top quark-antiquark pair,
which could arise through the mediation of very heavy new states.
In Eq.~(\ref{eq:lag}), \mt is the mass of the top quark, and $\text{G}^{\mu\nu}$ the gluon field.
We neglect possible modifications of the bottom quark Yukawa coupling \kappab, which is already constrained by LHC data~\cite{Corbett:2015ksa}.
The Feynman diagrams contributing to \HH production in \pp collisions at leading order (LO) are shown in Fig.~\ref{fig:dia}.
The translation of the above parametrization to the flavour-diagonal Higgs basis (as discussed in Ref.~\cite{FALKOWSKI2016}) is trivial;
we use the notation of Eq.~(\ref{eq:lag}) for simplicity.

The parameter space for the Higgs boson couplings in a BSM scenario has five dimensions.
Constraints on the ranges of the five parameters come from measurements of single Higgs boson production already performed at the LHC,
as well as other theoretical considerations~\cite{Carvalho:2015ttv}.
However, the allowed phase space is still large and a precise scan is not feasible.
The kinematical properties of the pair-produced Higgs bosons depend strongly on the values of those five couplings.
A statistical approach has been developed in order to identify regions of the parameter space that have similar final state kinematical properties.
The approach uses the generator-level distributions of the invariant mass of the HH system (\mHHgen) and the modulus of the cosine of the polar angle of one Higgs boson with respect to the beam axis ($\abs{\ctstarGen}$) to cluster the points in the parameter space.
In each region found, a representative benchmark model is selected as the point having most similar kinematical properties to the other points in the region.
The procedure, described in Ref.~\cite{Carvalho:2015ttv}, leads to twelve benchmarks that taken together best represent, within a limited uncertainty,
the phenomenology of the whole five-dimensional space.
An additional benchmark (box), representative of the null Higgs boson self-coupling hypothesis, is also considered.
The parameter values of the benchmarks are listed in Table~\ref{table:benchmarks}. The search for BSM signals presented here is focused on these benchmarks.

\begin{table}[htbp]
\topcaption{The values of the anomalous coupling parameters for the 13 benchmark models studied~\cite{Carvalho:2015ttv}.
For reference, the values of the parameters in the SM are also included. }
 \centering
 \begin{tabular}{l c c c c c }
\hline
Benchmark point & \kappalambda & \kappat & \ctwo & \cg & \ctwog \\
\hline
1 & 7.5 & 1.0 & -1.0 & 0.0 & 0.0 \\
2  & 1.0 & 1.0 & 0.5 & -0.8 & 0.6\\
3 & 1.0 & 1.0 & -1.5 & 0.0 & -0.8\\
4 & -3.5 & 1.5 & -3.0 & 0.0 & 0.0\\
5  & 1.0 & 1.0 & 0.0 & 0.8 & -1.0\\
6  & 2.4 & 1.0 & 0.0 & 0.2 & -0.2\\
7 & 5.0 & 1.0 & 0.0 & 0.2 & -0.2\\
8 & 15.0 & 1.0 & 0.0 & -1.0 & 1.0\\
9 & 1.0 & 1.0 & 1.0 & -0.6 & 0.6  \\
10 & 10.0 & 1.5 & -1.0 & 0.0 & 0.0\\
11  & 2.4 & 1.0 & 0.0 & 1.0 & -1.0\\
12 & 15.0 & 1.0 & 1.0 & 0.0 & 0.0 \\
Box & 0.0 & 1.0 & 0.0 & 0.0 & 0.0 \\ [\cmsTabSkip]
SM & 1.0 & 1.0 & 0.0 & 0.0 & 0.0 \\
\hline
\end{tabular}
\label{table:benchmarks}
\end{table}

\section{The CMS detector}
The CMS detector is a multipurpose apparatus designed to reconstruct
the high-energy interactions produced by the LHC.
Its central feature is a superconducting solenoid with an internal
diameter of 6\unit{m}. The solenoid generates a magnetic field of 3.8\unit{T} inside a
volume occupied by four main sub-detectors, each composed of a barrel and two endcap sections:
silicon pixel and strip tracker detectors,
a lead tungstate crystal electromagnetic calorimeter (ECAL), and a
brass and scintillator hadron calorimeter (HCAL).
The pixel tracker provides an impact parameter
resolution for charged tracks of about 15\mum,
which allows for a precise reconstruction of secondary vertices,
crucially used to identify jets originating from the hadronization of \cPqb\ quarks (\cPqb\ jets).
Muons are measured in gas-ionization detectors embedded in
a steel flux return yoke outside the solenoid.
Information from the calorimeters and muon detectors is used by the first level of
the CMS trigger~\cite{Khachatryan:2016bia}, a system based on custom hardware processors
that provides the first online event selection. The second
level of the CMS trigger, also called high-level trigger and
consisting of a farm of processors running a version of the full event reconstruction software optimized for fast processing,
further selects events using information from the whole detector
before sending them downstream for detailed processing and storage.
Particles produced in the \pp collisions are detected in the
pseudorapidity range $\abs{\eta} < 5$.
Pseudorapidity is defined as $\abs{\eta} = -\ln \tan(\theta/2)$, where the polar angle $\theta$ is measured from the z-axis,
which points along the beam direction toward the Jura mountains from LHC Point 5.
A more detailed description of the CMS detector
can be found in~\cite{Chatrchyan:2008zzk}.

\section{Data sets}
\label{sec:samples} \label{s:datasets}
The online event selection for the data used in this analysis is designed to select a sample of multijet events enriched with \cPqb\ quark decays,
reducing the rate from the QCD multijet background with light quarks and gluons.
The combined secondary vertex (CSVv2) algorithm~\cite{Sirunyan:2017ezt} is used
to identify \cPqb\ jets. This algorithm exploits the
relatively long lifetime of hadrons containing \cPqb\ quarks ($c\tau \sim 450\mum$), which results in
a displaced decay point of the produced \cPqb\ hadrons.
The reconstructed trajectories of charged decay products from \cPqb\ hadrons thus exhibit
significant impact parameters with respect to the \cPqb\ quark production point.
The CSVv2 algorithm uses the impact parameter information together with information on other characteristics of the jets
to discriminate jets originating from \cPqb\ quarks from those produced by the hadronization of light quarks or gluons.
Two trigger paths contribute to the online selection.
In the first trigger path jets are considered if their momentum transverse to the beam direction, \pt, is above 30\GeV and $\abs{\eta} < 2.6$.
Selected events must contain at least four such jets of which at least three are tagged as \cPqb\ jets by the CSVv2 algorithm
and at least two have $\pt > 90\GeV$.
The second trigger path requires at least four jets with $\pt > 45\GeV$ with at least three tagged as \cPqb\ jets by the CSVv2 algorithm.
The logical \textsc{or} between these two selections provides
the data used in this analysis.

{\tolerance=800
The production of nonresonant \HH in the SM
is simulated following the prescriptions of Ref.~\cite{Hespel:2014sla} at LO with \MGvATNLO 2.2.2~\cite{Alwall:2014hca} used as the generator.
Loop factors are calculated on an event-by-event basis and applied to an effective model,
from Ref.~\cite{Hespel:2014sla}; the NNPDF30\_lo\_as\_0130\_nf\_4 PDF set~\cite{nnpdf} is used.
In addition, for the study of BSM models involving anomalous Higgs boson couplings,
we generate for each of the parameter space points listed in Table~\ref{table:benchmarks} a set of 300\,000 simulated events.\par}

The 14 simulated signal samples are added together to obtain a larger signal sample. We will refer to this ensemble of events as the Pangea sample.
This sample is then reweighted to reproduce the physics of any particular point in the BSM phase space.
The weights are obtained by looking at the matrix element information for \mHHgen and \ctstarGen from dedicated simulations, as described in Ref.~\cite{carvalho2017}.
The numbers of events used to determine the weights at generator level are 3\,000\,000 for the SM sample and 50\,000 for each BSM benchmark.
In the following, we always use the Pangea sample instead of the 14 original samples to study signal properties in each model considered.

{\tolerance=800
Although our search employs an approach fully based on data to model backgrounds,
we make use of a simulation of QCD processes for several cross-checks.
This simulation consists of a collection of seven simulated data sets of contiguous ranges in the \HTgen variable,
which is defined as the scalar sum of the \pt of all partons that originate from the hard-scattering process in a simulated event.
The samples are generated by \MGvATNLO 2.2.2 at LO, using the NNPDF30\_lo\_as\_0130 set, and are then interfaced with \PYTHIA 8.212~\cite{Sjostrand:2014zea}
for fragmentation and parton showering, using the MLM matching~\cite{Alwall:2007fs};
their equivalent integrated luminosity
depends on the \HTgen range considered and increases from 0.06 to 400\fbinv as \HTgen varies between 200 and 2000\GeV.
For additional studies of the sub-dominant background from top quark pairs, a large
next-to-leading order (NLO) \POWHEG 2.0~\cite{Nason:2004rx,Frixione:2007vw,Alioli:2010xd} sample of inclusive \ttbar~\cite{Campbell:2014kua} events is used.
The behaviour of minor backgrounds is verified and a study of their contamination of our selected sample
is carried out using \POWHEG 2.0 NLO samples of single top quark $t$ channel~\cite{Alioli:2009je},
\ttH~\cite{Hartanto:2015uka}, single Higgs boson production~\cite{Bagnaschi:2011tu}, and associated \ZH production~\cite{Luisoni:2013kna}.
In addition, we use single top quark $s$ channel, \tttt,
\ttbb, and \bbH samples generated with \MGvATNLO 2.2.2 at NLO.
All of those samples are interfaced with \PYTHIA 8.212 for parton showering and fragmentation.
The \ttbar sample utilises the generator tune {CUETP8M2T4}~\cite{CMS-PAS-TOP-16-021} for the underlying event activity,
other samples interfaced with \PYTHIA use the tune {CUETP8M1}~\cite{Khachatryan:2015pea}.
The \ttbar, \ttH, single Higgs boson, and \ZH samples are generated using the  NNPDF30\_nlo\_as\_0118 PDF set.
The single top quark, \ttbb, and \bbH samples are generated with the NNPDF30\_nlo\_nf\_4\_pdfas set.
The NNPDF30\_nlo\_nf\_5\_pdfas set is used to generate the \tttt sample.
All of the PDF sets are taken from the LHAPDF6 set~\cite{Buckley:2014ana}.
The response of the CMS detector is modelled using
\GEANTfour~\cite{Agostinelli2003250}.\par}

Finally, in order to study possible discrepancies between the efficiency of the triggers used in our data selection and their modelling by the simulation,
we compare the effect of \cPqb\ jet selection requirements on data collected by a trigger requiring a single isolated muon of $\pt > 18\GeV$ with its simulation,
using a mixture of events from \ttbar/single-top described above and a \MGvATNLO 2.2.2 \PW +jets LO sample using the MLM matching,
weighted appropriately, and a \PW +jets sample generated using the NNPDF30\_lo\_as\_0130 PDF set.

\section{Event reconstruction}
\label{s:event_reco}
Global event reconstruction is performed by the particle-flow (PF)
algorithm~\cite{CMS-PRF-14-001}, which
aims to reconstruct and identify each individual particle in an event,
with an optimized combination of information from the various elements of the CMS detector.
In this process, the identification of the particle type (photon, electron, muon, charged hadron, neutral hadron) plays an important role in the determination of the particle direction and energy.
Electrons (\eg coming from photon conversions in the tracker material or from \cPqb\ hadron semileptonic decays) are identified as a primary charged particle track and one or more ECAL energy clusters corresponding to this track extrapolation to the ECAL and to possible bremsstrahlung photons emitted along the way through the tracker material.
Muons (\eg from \cPqb\ hadron semileptonic decays) are identified as a track in the central tracker consistent with either a track or several hits in the muon system, associated with an energy deficit in the calorimeters.
The objects primarily considered in this analysis are hadronic jets,
composed of particles produced by quark fragmentation and hadronization.
The energy of charged hadrons is determined from a combination of their momentum measured in the tracker and the matching ECAL and HCAL energy deposits, corrected for zero-suppression effects and for the response function of the calorimeters to hadronic showers.
The energy of neutral hadrons is obtained from the corresponding corrected ECAL and HCAL energy.
Jets are reconstructed from PF candidates using the anti-\kt
clustering algorithm~\cite{Cacciari:2008gp}
with a distance parameter of 0.4, as implemented in the \FASTJET
package~\cite{Cacciari:2011ma}.
Jet energy corrections are applied to both data and simulation
to scale the energy and correct for differences in the detector response in real and simulated collisions~\cite{Khachatryan:2016kdb}.
Jet identification criteria are also applied in order to reject fake jets
from detector noise and jets originating from primary vertices
not associated with the hard interaction~\cite{CMS-PAS-JME-13-005}.
The combined multivariate algorithm (cMVAv2)~\cite{Sirunyan:2017ezt}
is used in the offline analysis to identify jets that originate
from the hadronization of \cPqb\ quarks.
The cMVAv2 builds on the CSV algorithm by adding soft-lepton information to the combined discriminant.
The value of the multivariate discriminant used depends on the required suppression of jets from light quarks and gluons.
The medium working point of the cMVAv2, defined such that the misidentification
rate of light quarks and gluons as \cPqb\ jets is 1\%, is used in this analysis. For jets produced by the hadronization of \cPqb\ quarks emitted in \HH
production events, the medium working point corresponds to a \cPqb-tagging efficiency of about 65\% for the jets of interest of this analysis.

A weight is applied to each Monte Carlo (MC) event in order to match
the distribution of the number of primary interactions per event
in data (pileup correction), thus reproducing the effect on the selection efficiency of the
varying instantaneous luminosity conditions incurred during data taking.
The simulated events are also weighted to account for measured differences
in the \cPqb\ tagging efficiency between data and simulation~\cite{Sirunyan:2017ezt}.
The trigger efficiency for signal events is evaluated using a full simulation of the trigger~\cite{Khachatryan:2016bia}.
The correction factor for the efficiency is found to be $0.96 \pm 0.02$ based on measurements performed in \cPqb-tag multiplicity categories, using a top-pair enriched sample collected with an inclusive muon trigger.

\section{Analysis strategy}
\label{sec:anStr}
The focus of this search is the study of nonresonant production of \HH in the \bbbb  final state, as predicted by the SM and by several BSM extensions.
The analysis is optimized for sensitivity to the SM signal. We use the same selection to extract limits on the \HH production cross section for the BSM models.

The offline selection, performed on all data events passing one of the two trigger paths described
in Section~\ref{s:datasets}, aims at increasing the fraction of data events containing
two Higgs boson candidates decaying into \cPqb\ quark jet pairs. This includes a preliminary selection of events where
four or more jets have been \cPqb-tagged by the cMVAv2.
Although this selection significantly reduces the QCD multijet background rate, this background still dominates the selected data,
with contributions from events where light quark or gluon jets are mistagged by the cMVAv2 and events containing heavy quarks.

After the selection of events with four or more \cPqb\ tags, each reconstructed Higgs boson candidate is composed of a pair of \cPqb\ jets,
referred to in the following as ``a dijet system'' or simply ``dijet''. A boosted decision tree (BDT) classifier \cite{xgboost}
is then trained to exploit the observable differences between the SM signal and the background.
Finally, a search for a signal contribution to the selected data and an extraction of an upper limit in the number of selected signal events
is performed by means of a binned fit to the distribution of the BDT classifier output.
The limit on the number of events is converted to a limit on the \HH production cross section times the square of the branching fraction of
the Higgs boson into a \bbbar pair, using the corresponding integrated luminosity and the computed signal efficiency.

Both the optimization of the BDT classifier and the extraction of upper limits on \HH production require an accurate modelling of the multijet background.
Unfortunately, the precise simulation of QCD processes yielding a large number of final-state partons is notoriously hard,
as MC simulations are not complete to beyond LO;
in addition, the very large production cross section for those processes makes it wholly impractical to produce simulated data sets
corresponding to an integrated luminosity comparable to that of collision data.
To address these issues, a dedicated method, fully based on data, was developed to produce a precise model of the kinematical behaviour of background events.
This is described in detail in Section~\ref{sec:background}.

\section{Event selection}
\label{sec:evtSel}
The events of interest are identified by a jet-based selection applied to data collected by the triggers described in Section~\ref{sec:samples},
as well as to all simulated samples.
Jets are required to have $\ptJet > 30\GeV$ and $\absEtaJet < 2.4$.
We require at least four such jets ($\Nj \geq 4$)
and these need to be defined as \cPqb-tagged jets by the medium working point of the cMVAv2 ($\Nb \geq 4$).
These criteria strongly reduce the QCD multijet background and select \HH production events where the final state can be fully reconstructed.
The number of selected events in the data set studied is 184\,879.

The efficiencies for the SM signal are listed in Table~\ref{tab:sigEff}.
The efficiency for the 13 BSM points varies from -40\% to +10\% compared to the SM values.
The average number of jets per selected event is ${\approx}5$.
The four jets with the highest cMVAv2 discriminant values are considered as the decay products of two Higgs boson candidates.
The pairing of the four jets into Higgs boson candidates is performed by considering the invariant mass of the two dijet candidates calculable for
the three possible pairings, and computing the absolute mass differences $\deltaMklmn = \diffMklmn$, where the $klmn$ indices
run on the three permutations of the four jets.
The combination resulting in the smallest mass difference between the two dijet systems
is chosen as the one best describing the decay topology.
This procedure results in a correct pairing of the \cPqb\ quarks to Higgs bosons in 54\% of the cases, as tested on the Pangea MC signal sample.
The two selected dijets are then labelled as ``leading'' and ``trailing'' according to their invariant mass value.
This procedure, which does not explicitly use the known mass of the Higgs boson,
allows the dijet masses for the selected combinations to retain the power to discriminate the \HH production signal from the background.

\begin{table}[htb]
 \topcaption{Cut-flow efficiency for the SM signal \ppToHHbbbb;
 the efficiency and the relative reduction of each successive selection step is shown.
 The number of expected SM signal events for an integrated luminosity of 1\fbinv is also reported. }
\centering
 \small
\begin{tabular}{r l l  c c}
  \hline
                             & Produced  & Trigger & $\geq$4 \cPqb\ tags \\
  \hline
   N events / fb     & 11.4     & 3.9   & 0.22 \\
   Relative eff.              &          & 34\%  & 5.6\% \\
   Efficiency                 &          & 34\%  & 1.9\% \\
  \hline
 \end{tabular}
 \label{tab:sigEff}
\end{table}

A multivariate technique is used in order to improve the sensitivity of the analysis.
A BDT discriminator is trained to distinguish the SM signal from backgrounds (as described in Section~\ref{sec:background}),
using the \textsc{XGBoost} library~\cite{xgboost}.
All 13 BSM models use the same BDT as the SM to distinguish signal from background.
We supply the BDT algorithm with a set of variables describing the kinematical properties of the event.
The list of variables is pruned to discard those not contributing to the overall discrimination power of the algorithm.
In Table~\ref{tab:mvaVars} we list the variables chosen to build the classifier.
In order to characterise the \HH system we use as mass variables the invariant mass of the \HH system (\MHH),
an estimator of the combined mass of the \HH system, \MX\ (defined by Eq.~\ref{eq:mx} below),
and the invariant masses of the reconstructed Higgs boson candidates (\MHone, \MHtwo).
The \pt of the \HH system and of each Higgs boson candidate
 (\pTHH, \pTHone\ and \pTHtwo) are used as well as the
$\DR = \sqrt{\smash[b]{(\Deta)^2 + (\DPhi)^2}}$ and \DPhi\ angles between the jets that form each reconstructed Higgs boson
(\DRjjHone, \DRjjHtwo, \DPhijjHone and \DPhijjHtwo).
Additionally, we use the \thetastar\ angles between the \HH system and the leading Higgs boson candidate, \ctsHoto,
and between the leading Higgs boson candidate and the leading jet, \ctsHoneJone.
We further use the following jet-related variables: the \pTjets and \etaJets ($i = 1\text{--}4$) of
the four jets with the highest values of the cMVAv2 discriminant, the scalar \pt sum of the jets in the event (\HT),
and of the jets that are not part of the reconstructed \HH system (i.e. the rest of the jets, \HTrest).
The cMVAv2 values of the third and fourth jets sorted by cMVAv2 value ($\cmva{3}$ and $\cmva{4}$) are also used.
The estimator, \MX, of the mass of the system of two Higgs bosons is constructed as:
\begin{linenomath}
\begin{equation}
\MX = \Mfourj - \Big(\MHone-\mHiggs\Big) - \Big(\MHtwo-\mHiggs\Big),
\label{eq:mx}
\end{equation}
\end{linenomath}
where $\mHiggs = 125\GeV$. Even though \MX\ is strongly correlated with other variables used in the BDT, its use improves the discrimination power.

The invariant masses of the reconstructed Higgs boson candidates are the variables with the largest discrimination power,
but all the variables used make significant contributions to the classifier.

\begin{table}[htbp]
 \topcaption{List of BDT input variables.}
 \centering
 \begin{tabular}{l l l}
   \hline
   \HH system     & \PH candidates      & Jet variables \\
   \hline
   \MX, \MHH,     & \MHone,  \MHtwo   & \pTjetsFour, \etaJetsFour,  \\
   \pTHH          & \pTHone, \pTHtwo  & \HTrest, \HT \\
   \ctsHoto       & \ctsHoneJone        & $\cmva{3}$, $\cmva{4}$ \\
                  & \DRjjHone, \DRjjHtwo,  \DPhijjHone, \DPhijjHtwo & \\
   \hline
 \end{tabular}
 \label{tab:mvaVars}
\end{table}
We use 60\% of the Pangea sample for training the classifier.
The remaining 40\% of the Pangea sample is employed for the validation (20\%) and application (20\%) steps;
this splitting has been found to produce maximal sensitivity to a possible \HH signal.
As a background sample, an artificial data set constructed with a custom mixing procedure, as described in Section~\ref{sec:background}, is employed.

\section{The background model}
\label{sec:background}
A method exploiting collision data only, based on hemisphere mixing, has been developed~\cite{hem_mixing} to perform two separate tasks:
first, to provide input to the training of the BDT classifier;
and second, to reproduce the expected shape of the BDT output in background-only events.
The method does not require the presence of signal-depleted sidebands in order to extract a background estimation;
in fact, it aims at creating an artificial background data set using the whole original data set as the input.
Thus, rather than a model of a single distribution, a full model of the original data is produced.

\subsection{The hemisphere mixing technique}
\begin{figure}[htbp]
\centering
\includegraphics[width=1.0\textwidth]{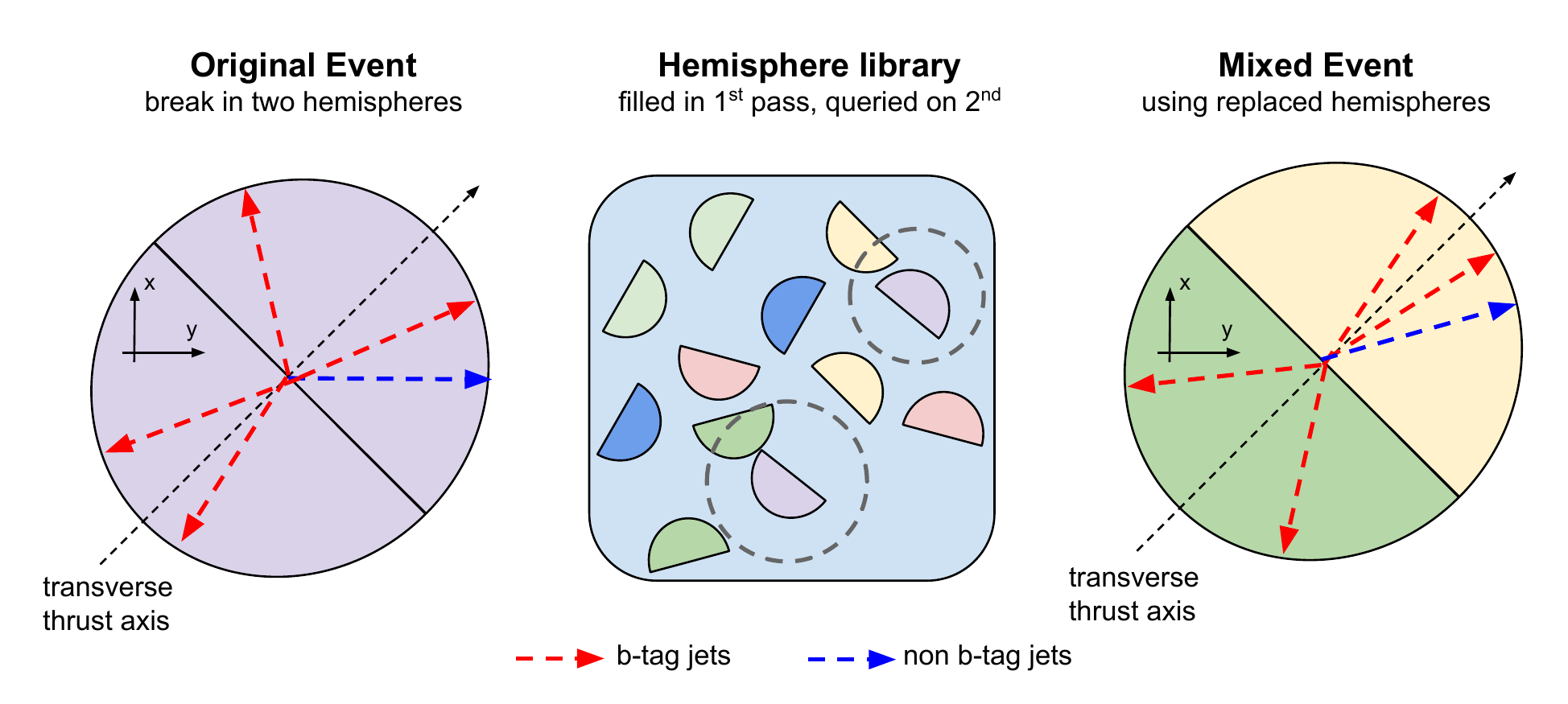}
\caption {An illustration of the hemisphere mixing procedure.
The transverse thrust axis is defined as the axis on which the sum of the absolute values of the projections of the \pt of the jets is maximal.
Once the thrust axis is identified, the event is divided into two halves by cutting along
the axis perpendicular to the transverse thrust axis. One such half is called a hemisphere ($h$).
In a preliminary step, each event in the original $N$-event data set is split into two hemispheres that are collected in a library of $2N$ hemispheres.
Once the library is created, each event is used as a basis for creating artificial events.
These are constructed by picking two hemispheres from the library that are similar to the two hemispheres that make up the original event.}
\label{fig:hemisphere_mixing}
\end{figure}

The basic concept at the heart of the method is to divide each data event in two hemispheres.
The collection of hemispheres can then be used to create new events by recombining them in pairs.
To create a good background model, the kinematical properties of the new events must be as similar as possible to the
ones of the original data but also insensitive to the possible presence of signal.
In order to define the hemispheres, we use the transverse thrust axis.
This is defined as the axis on which the sum of the absolute values of the projections of the \pt of the jets is maximal, and correspondingly,
transverse thrust (\Thrust) is the value of this sum.
Once the transverse thrust axis is identified, the event is divided into two halves by cutting perpendicular to the transverse thrust axis.
One such half is called a hemisphere ($h$).
In a preliminary step, each event in the original $N$-event data set is split into two hemispheres that are collected in a library of $2N$ elements.
Once the library is created, each event is used as a basis for creating artificial events.
These are constructed by picking two hemispheres from the library that are {\em similar}, according to a measure defined below,
to the two hemispheres that make up the original event. An illustration of the procedure can be found in Fig.~\ref{fig:hemisphere_mixing}.

The number of jets \Njh\ and number of \cPqb-tagged jets \Nbh\ in each hemisphere,
together with four jet-related variables, are used to define a hemisphere similarity criterion.
The four variables are the combined invariant mass of all jets contained in the hemisphere \Mtoth,
transverse thrust of the hemisphere \Th,
the scalar sum of the projections of the \pt of all the jets onto the axis orthogonal to the thrust axis on the transverse plane, \Tah,
the projection of the vectorial sum of the momenta of the jets along the beam axis, \sumpzh.
If we label the original hemisphere $o$, and $q$ the one in the library that is compared to $o$,
the number of jets in $o$ and $q$ is required to be equal, $\Njind{o} = \Njind{q}$,
and also the number of \cPqb-tagged jets are required to be equal, $\Nbind{o} = \Nbind{q}$.
These two requirements are used to maintain the topology of the original events and
to avoid introducing events that would not pass the selection described in Section~\ref{sec:evtSel}
(\eg by combining a hemisphere with 2 jets with a hemisphere with 1 jet, resulting in an event with 3 jets).
The requirement for equal numbers of jets is waived for the infrequently occurring pairs of hemispheres
that both have at least four jets and at least four \cPqb-tagged jets.
For each hemisphere $q$ in the library fulfilling the above criteria, a multidimensional distance from hemisphere $o$ is computed using the four jet-related variables,
as follows:
\begin{linenomath}
\begin{equation}
\Dist{o}{q}^2 =
\frac{ (\Mtotind{o} - \Mtotind{q})^2} {V(\Mtot)}+
\frac{ (\Tind{o} - \Tind{q})^2}{V(T)}+
\frac{ (\Taind{o} - \Taind{q})^2} {V(\Ta)}+
\frac{ (\abs{\sumpzind{o}} - \abs{\sumpzind{q}})^2}{V(\sumpz)}.
\end{equation}
\end{linenomath}
In the equation above, $V(x)$ represents the variance for the variable $x$,
within the subset of events of given \Nb\ and \Nj\ characterizing the hemisphere in question.
Once all $\Dist{o}{q}$ are computed, the $k$th nearest-neighbour hemisphere in the library,
with $k \geq 1$ (\ie the one such that $0 = \Dist{o}{0} < \ldots < \Dist{q}{k}$) can be chosen to model the corresponding hemisphere of the original event;
the nearest hemisphere, corresponding to $k = 0$, is by construction the original one.
We match the \sumpz\ variables by considering only their absolute value
(assuming forward-backward symmetric detector acceptance to jets, as is safe to do in the case of the CMS detector)
and invert the sign of jet $\Sigma p_{z}$ components in one of the two matched hemispheres ($q_1$ and $q_2$) if
$\sgn(\sumpzind{o_1} \sumpzind{o_2}) \neq \sgn(\sumpzind{q_1} \sumpzind{q_2})$,
where indices $o_1$ and $o_2$ are the two hemispheres of the original event.
Finally, the four-vectors of the jets contained in the two hemispheres are rotated along the $\phi$ coordinate
to match the original transverse thrust axis of the modelled event.
To keep track of the distance criterion used to choose each hemisphere, the artificial event may be labelled as $(k_1,k_2)$ using the neighbour indices,
indicating that one hemisphere of the original event was replaced by its $k_{1}$th neighbour and the
other hemisphere of the original event by its $k_{2}$th neighbour and these were used to form the artificial event.
By applying this procedure to the whole set of events of the data to be modelled,
and by choosing a limiting value $K$ for $k$, we obtain a total of $K^2$ data sets, each equal in size to the original one,
and each featuring very similar characteristics to the original one, despite being made up entirely of artificial events.

The procedure described above is successful at modelling multijet events because it exploits the fact
that their production can be idealized at LO as a $2\to2$ process,
which is made complex by a number of sub-leading effects (QCD radiation, pileup, multiple interactions).
The reconstruction of the transverse thrust axis, and the decomposition of events into hemispheres using that axis as a seed,
uses the independent fragmentation of the two final state partons as a working hypothesis to create artificial replicas of the original events.
The method destroys any correlation in the jet distribution between the two hemispheres,
so that any physical effect, such as the decay of a heavy object into jet pairs, is washed out in the artificial samples.
Because of this, the resulting artificial data sets are unaffected by the presence of a small signal contamination in the original data.
This has been verified by signal injection tests.
We started with an original data set composed of simulated QCD multijet events to which is added an additional component of signal corresponding to a cross section 100  times larger than the one expected by the SM.
After hemisphere mixing, the kinematical properties of the resulting artificial samples are found to resemble closely those from the QCD multijet part of the original data set, which is its dominant component, and unaffected by the minority component (the signal contamination).
Naively this can be understood if we note that, if the signal fraction in the original sample amounts to \eg 0.1\%,
the probability that a signal event is modelled using two different hemispheres both originally belonging to signal events is of the order of 0.0001\%.
Event-based variables such as the two Higgs boson candidate masses, which are obtained by the minimum $\Delta M$ criterion described in Section~\ref{sec:evtSel}
and are thus sensitive to the characteristics of both hemispheres together,
do not retain their distribution in events where only one hemisphere is taken from an \HH decay event.

\begin{figure}[htbp]
\centering
\includegraphics[width=0.49\textwidth]{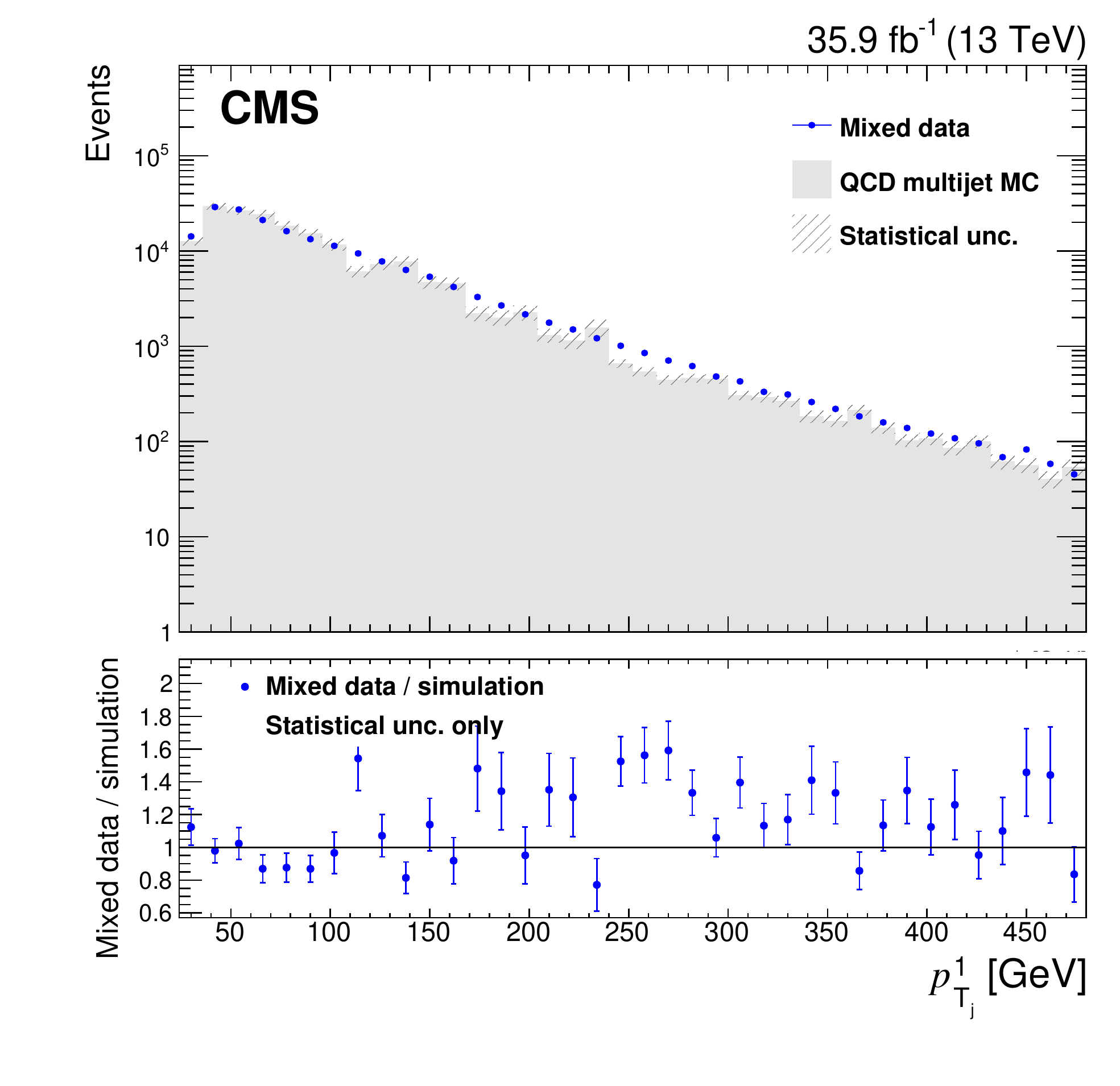}
\includegraphics[width=0.49\textwidth]{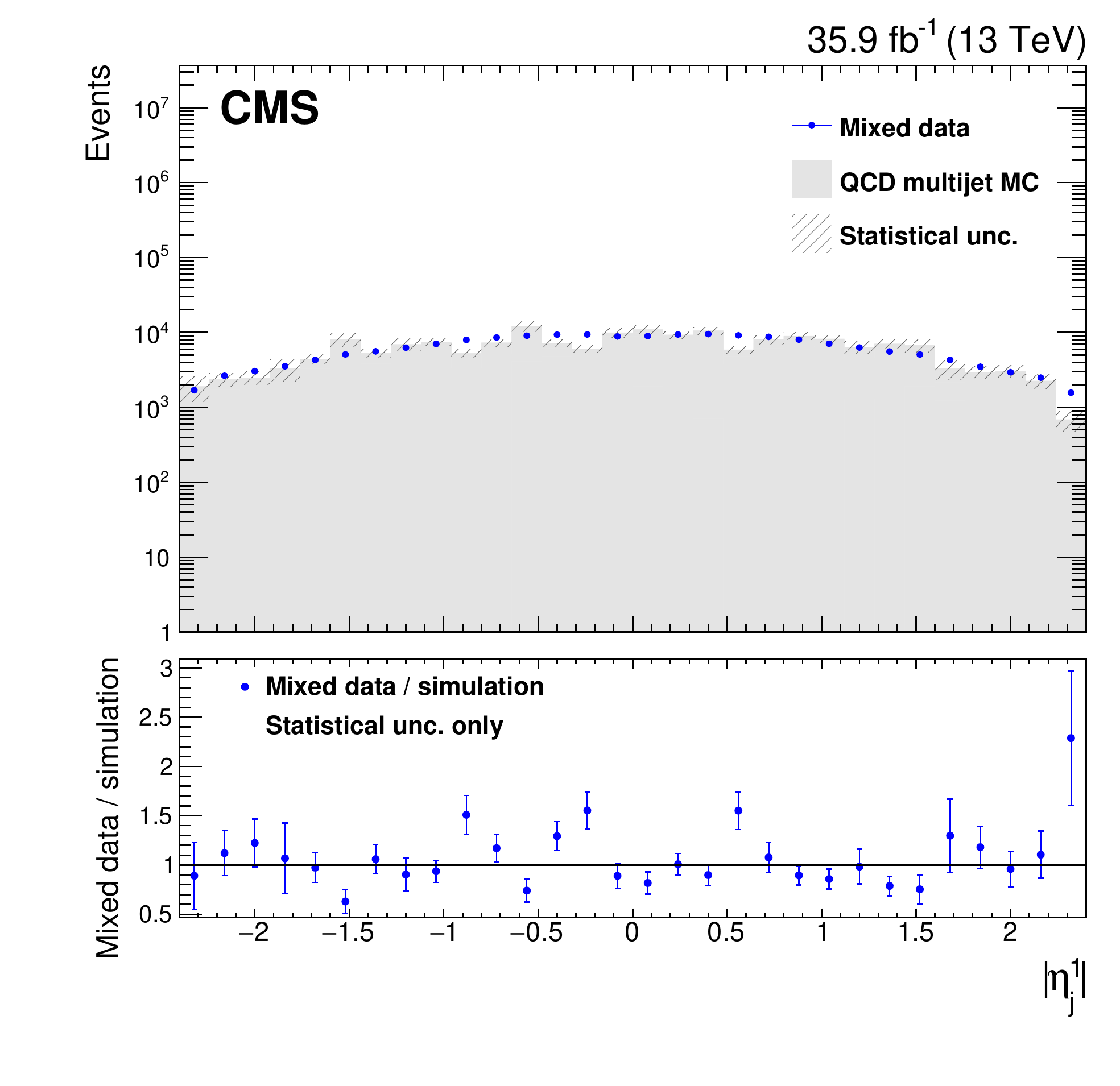}  \\
\includegraphics[width=0.49\textwidth]{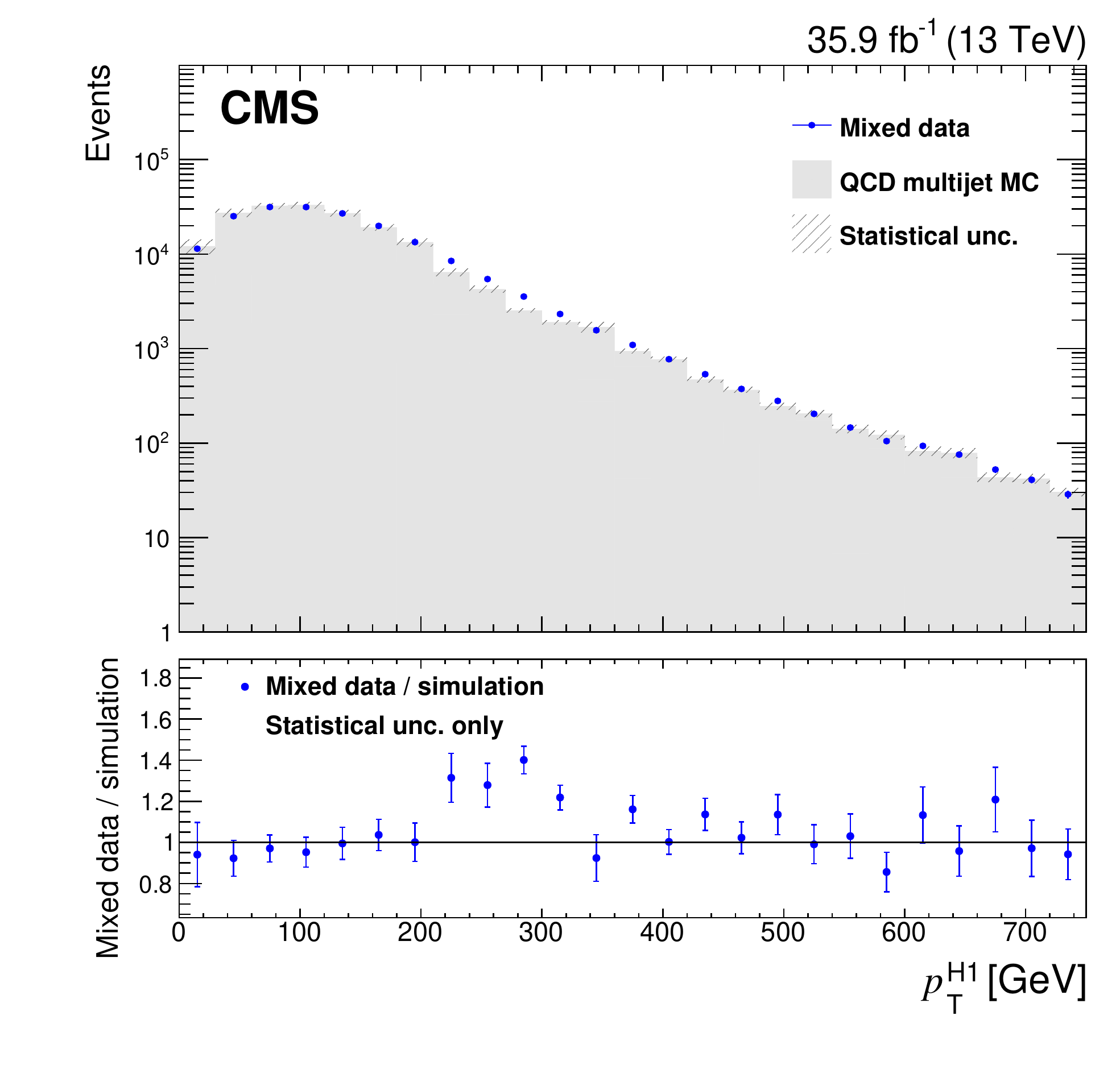}
\includegraphics[width=0.49\textwidth]{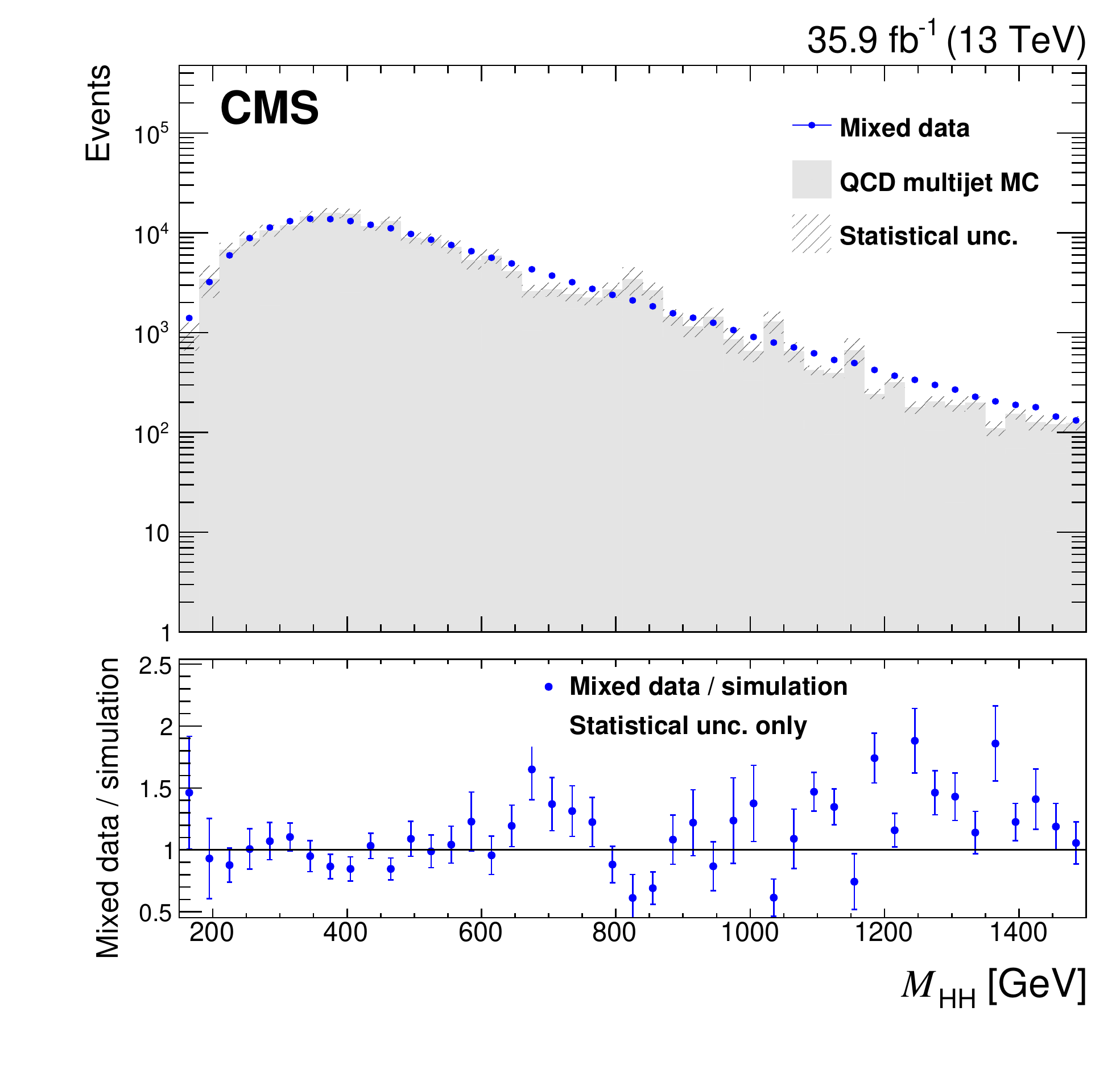}
\caption{Comparison between the background model obtained with the hemisphere mixing technique and MC simulation of QCD multijet processes for
$\pTjetind{1}$ (upper left), $\etaJetind{1}$ (upper right), \pTHone (lower left), and \MHH (lower right).
Bias correction for the background model, described in Section~\ref{sec:bias_corr}, is applied by rescaling the weight of each event using the event yield ratio between corrected and uncorrected BDT distributions. Only statistical uncertainties are shown as the uncertainties related to the bias correction can not be propagated from the BDT classifier
to a different variable.}
\label{fig:QCDMC}
\end{figure}

We apply the hemisphere mixing technique to data events selected with the $\Nj \geq 4$, $\Nb \geq 4$ criteria,
using $K = 10$ neighbour hemispheres to each hemisphere of the original event, which were found to still provide good modelling.
The resulting artificial samples are used to provide a background model in the training of the BDT classifier (training sample),
as well as an independent set for the BDT validation and optimization (validation sample),
and a third data set used to extract the predicted shape of the optimized BDT (application sample).
Not all of the data sets are fully independent so only a subset can be safely employed for further studies.
We use the following collections of artificial events in the measurement:
for the training sample, we use all artificial events of types $(1,1)$, $(1,2)$, $(2,1)$, and $(2,2)$;
for the validation sample, all artificial events of types $(3,4)$, $(5,6)$, $(7,8)$, and $(9,10)$; and for the application sample,
all artificial events of types $(4,3)$, $(6,5)$, $(8,7)$, and $(10,9)$.
This split guarantees that the three samples have equal number of events, and that the validation and application samples are independent of each other,
being constituted of artificial events made up of different hemispheres.
For the training sample the partial use of the same hemispheres in modelling different artificial events might at most slightly degrade the discrimination power
but does not have a detrimental effect on the subsequent steps of the analysis.
A study is performed by switching the validation and application samples and we find that this does not change the results.
The fraction of data events that are totally replicated in the background template is completely negligible.
A comparison between the distributions obtained through the procedure described above and by using MC simulation
for QCD multijet processes can be seen in Fig.~\ref{fig:QCDMC} for a number of variables.
The compatibility is good, although the statistical uncertainties in the model from MC simulation are large.

\subsection{The background template validation}
\label{sec:bias_corr}

\begin{figure}[htbp]
\centering
\includegraphics[width=0.49\textwidth]{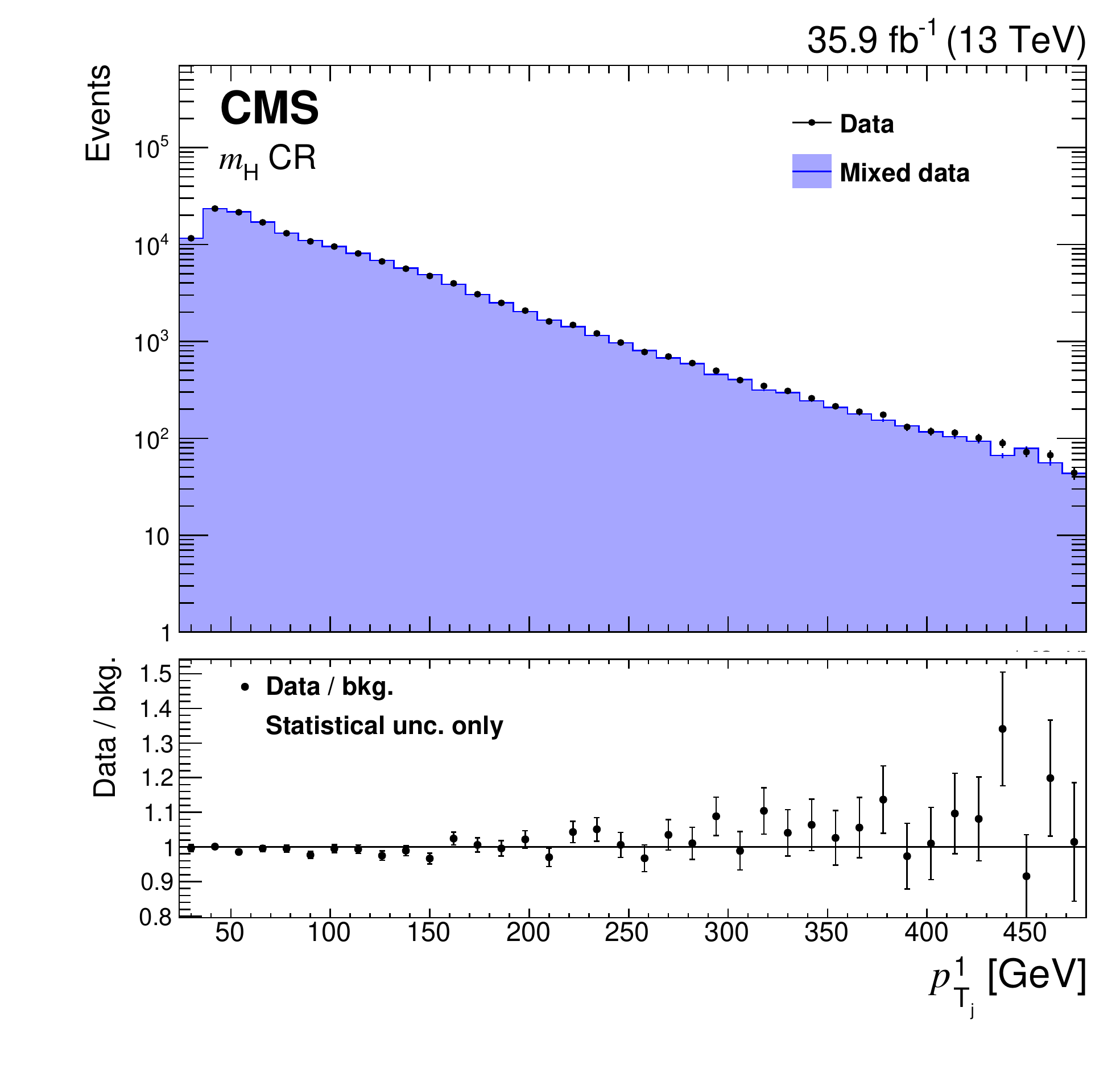}
\includegraphics[width=0.49\textwidth]{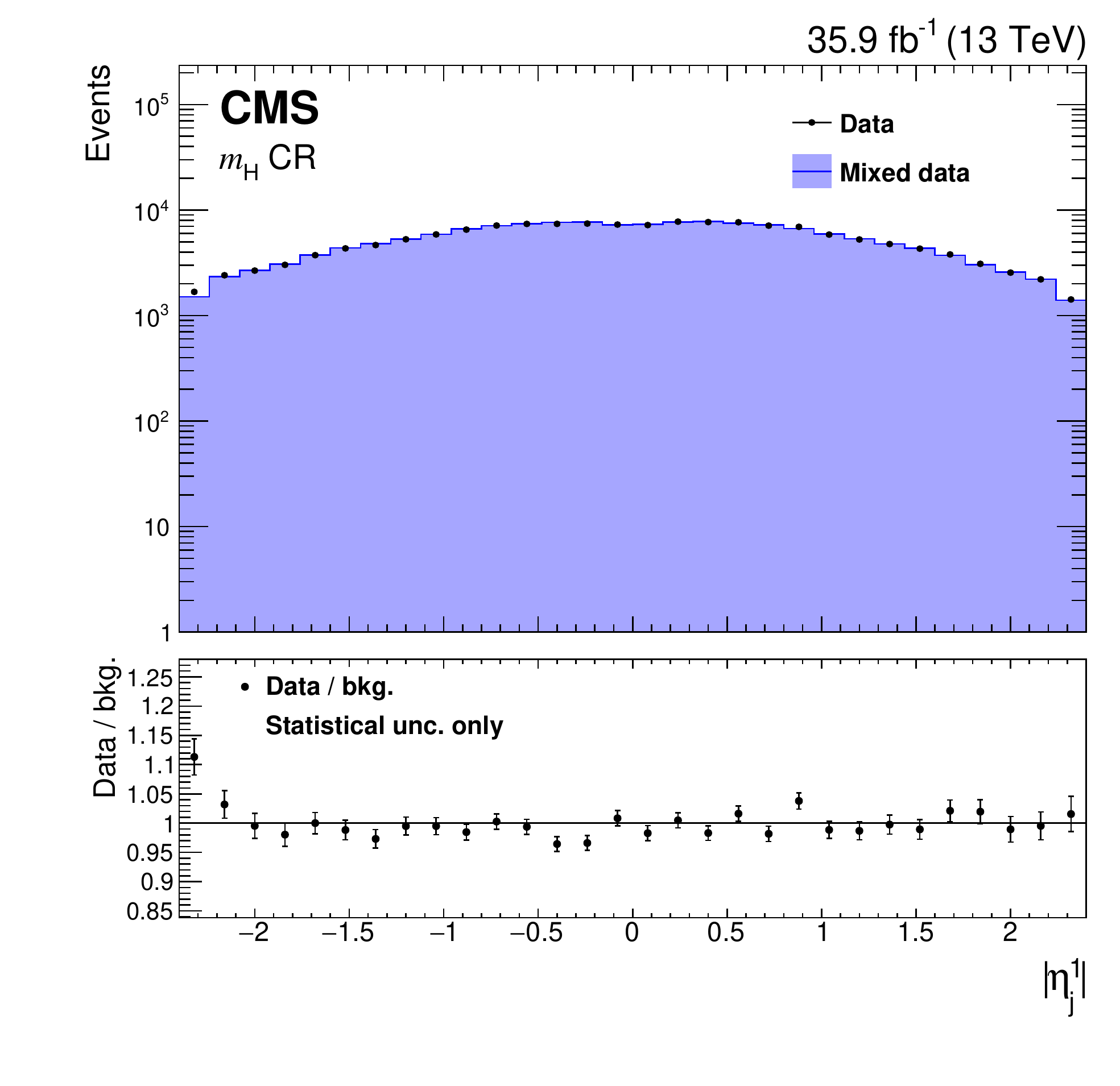}  \\
\includegraphics[width=0.49\textwidth]{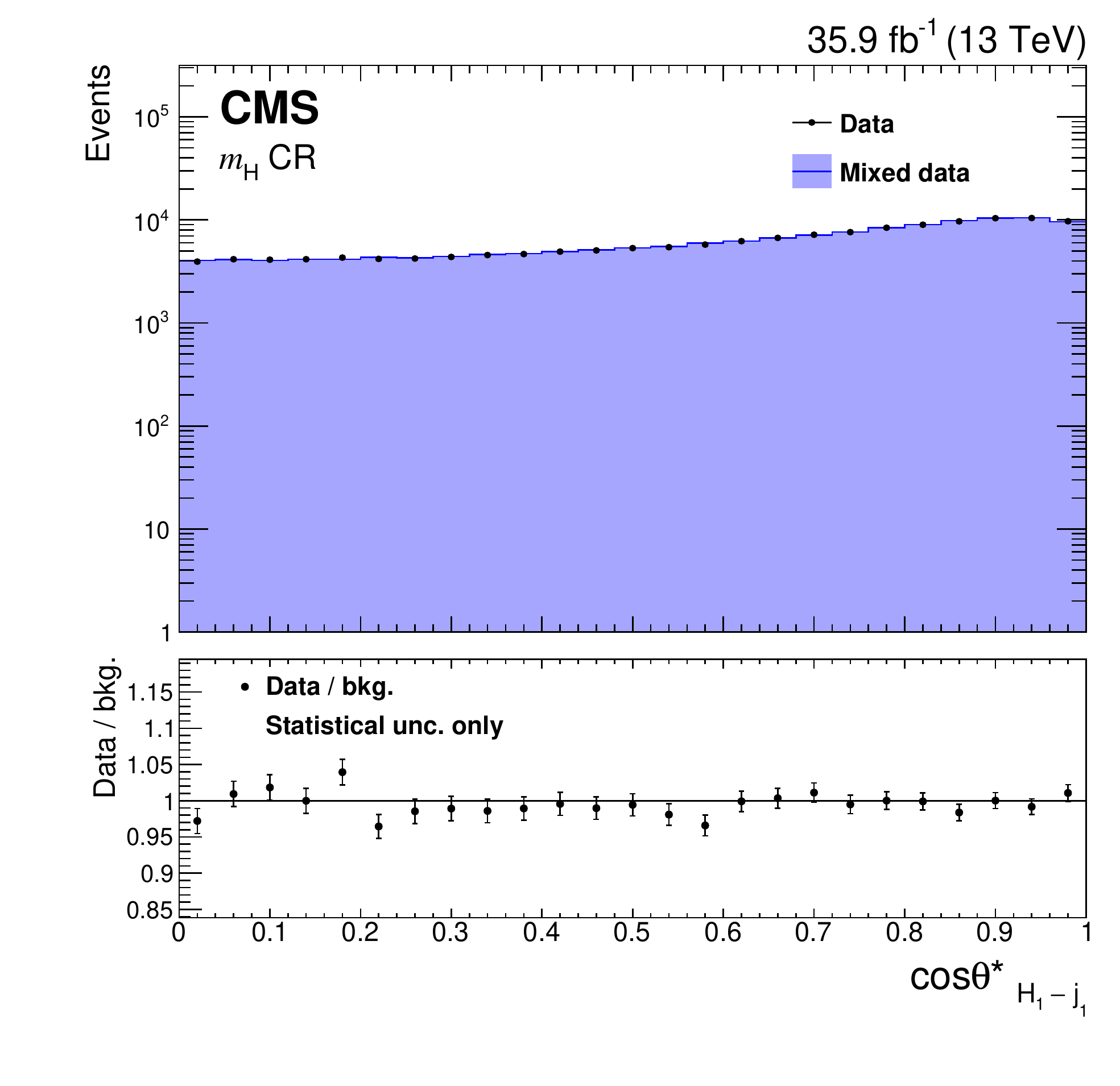}
\includegraphics[width=0.49\textwidth]{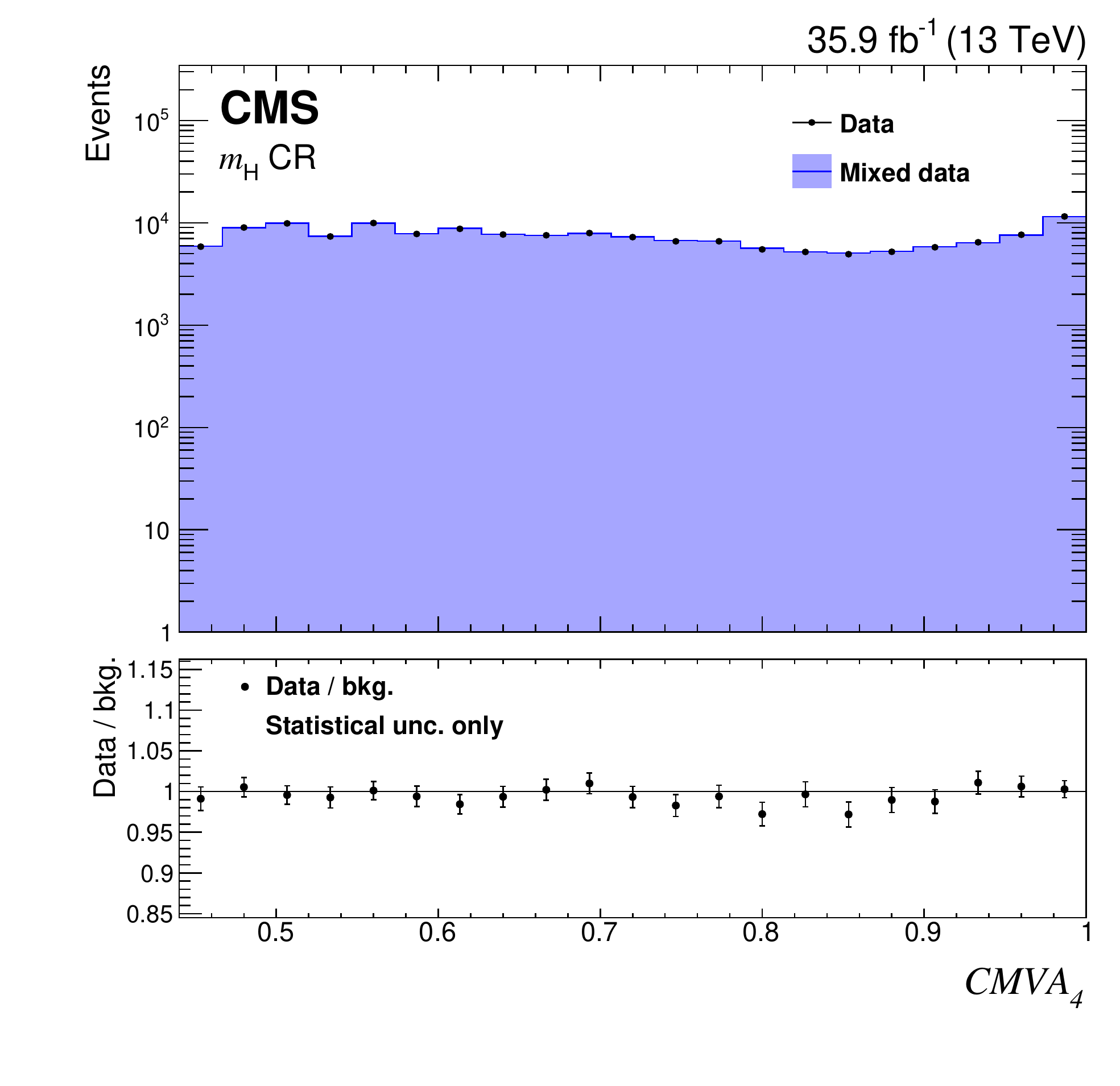}
\caption{Comparison between the background model obtained with the hemisphere mixing technique and data in the \mHiggs\ CR for
the variables $\pTjetind{1}$ (upper left), $\etaJetind{1}$ (upper right), \ctsHoneJone (lower left), and $\cmva{4}$ (lower right).
Bias correction for the background model, described in Section~\ref{sec:bias_corr}, is applied by rescaling the weight of each event using the event yield ratio between corrected and uncorrected BDT distributions in this CR.
Only statistical uncertainties are shown as the uncertainties related to the bias correction can not be propagated from the BDT classifier
to a different variable.}
\label{fig:massCR}
\end{figure}

\begin{figure}[htbp]
\centering
\includegraphics[width=0.49\textwidth]{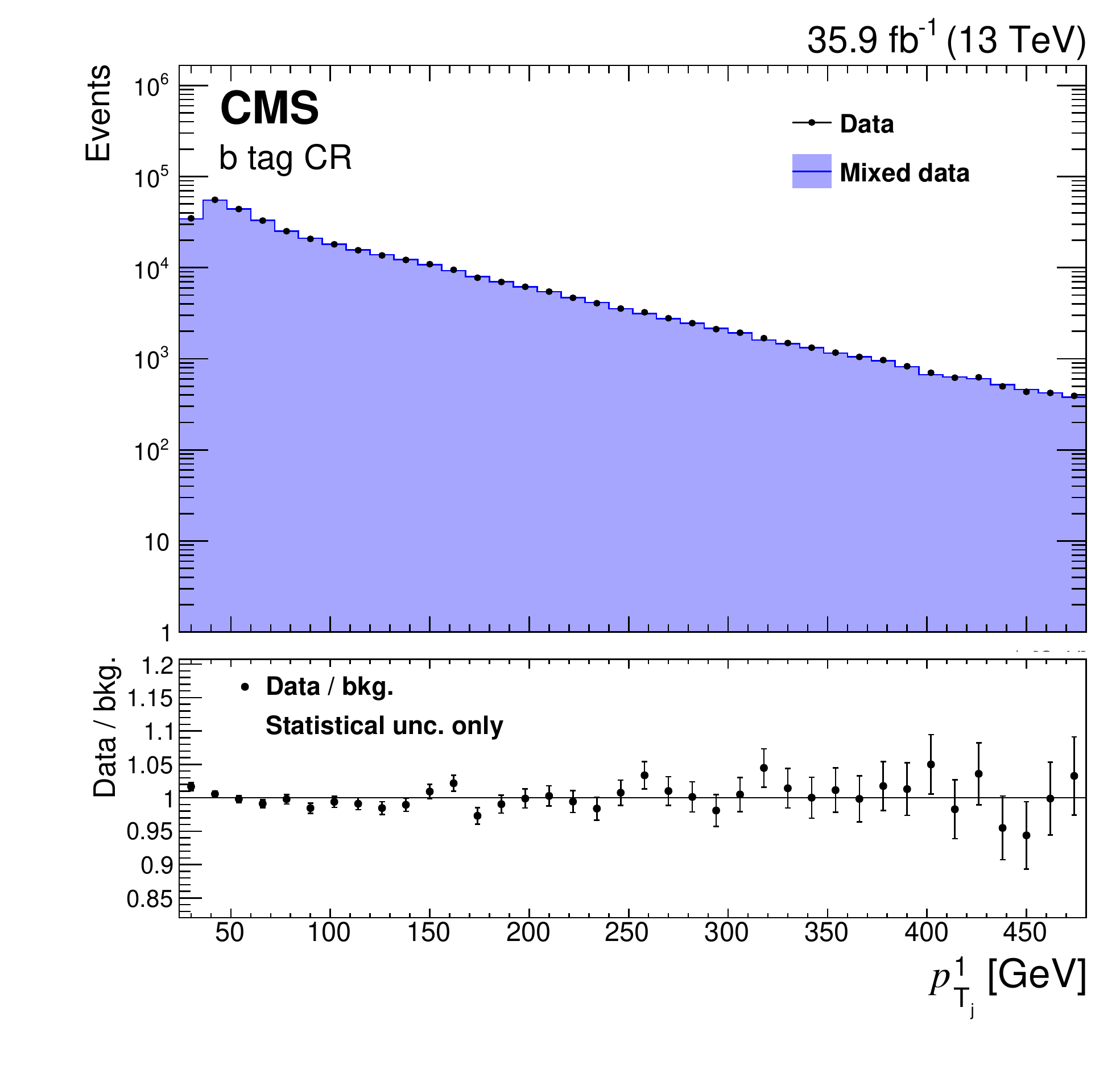}
\includegraphics[width=0.49\textwidth]{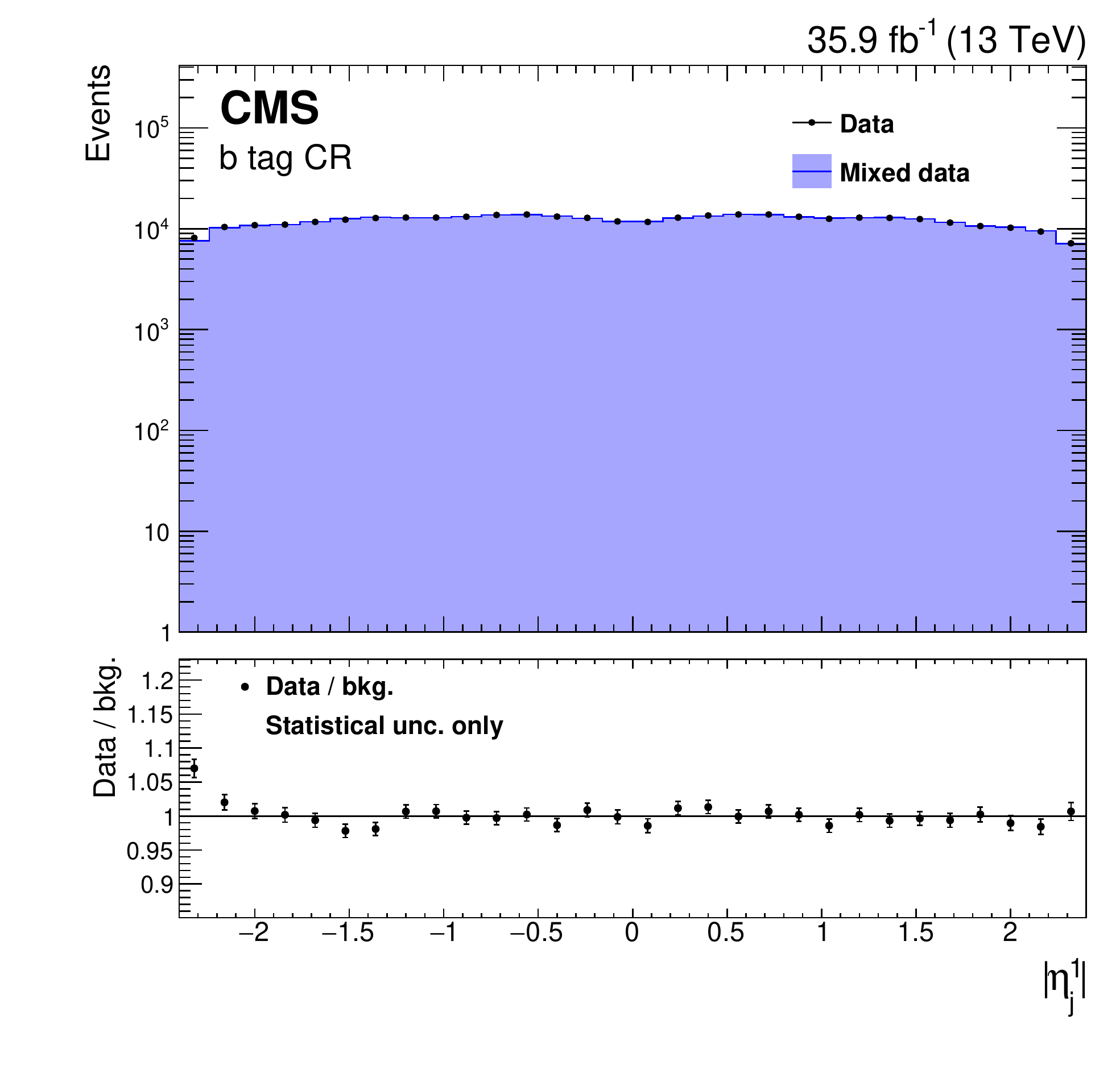}  \\
\includegraphics[width=0.49\textwidth]{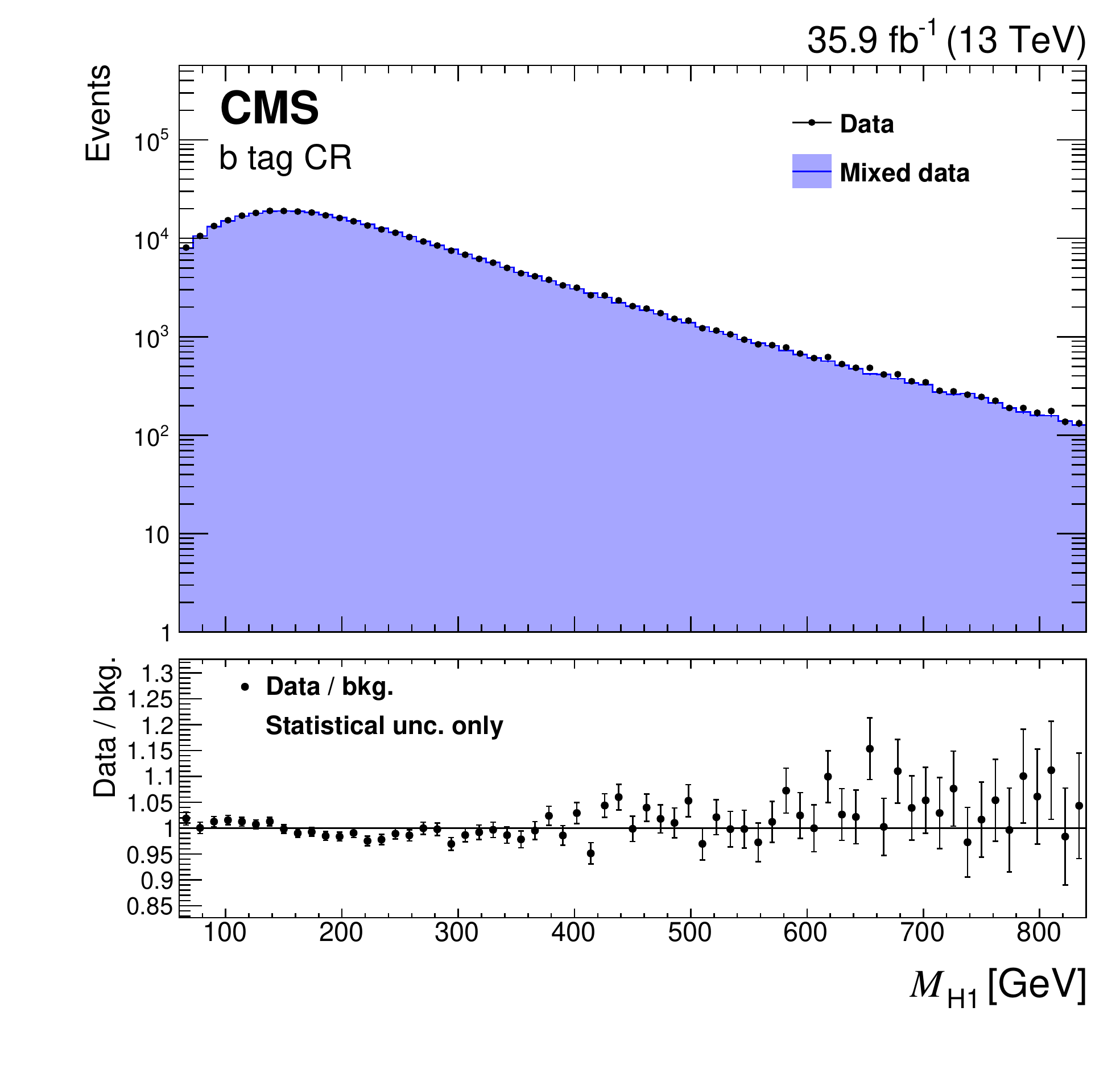}
\includegraphics[width=0.49\textwidth]{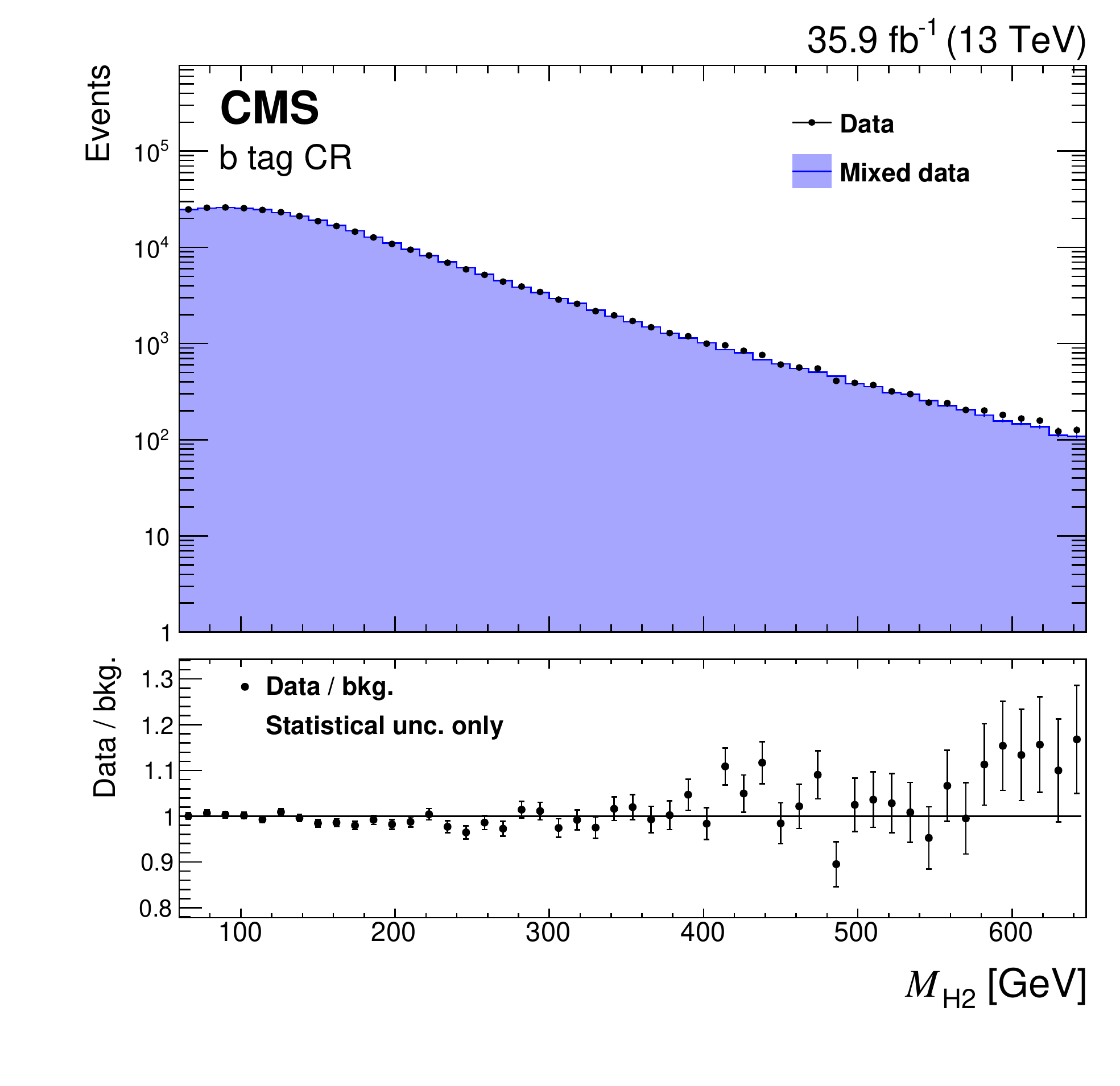}
\caption{Comparison between the background model obtained with the hemisphere mixing technique and data in the \cPqb\ tag CR for
the variables $\pTjetind{1}$ (upper left), $\etaJetind{1}$ (upper right), \MHone (lower left), and \MHtwo (lower right).
Bias correction for the background model, described in Section~\ref{sec:bias_corr}, is applied by rescaling the weight of each event using the event yield ratio between corrected and uncorrected BDT distributions in this CR.
Only statistical uncertainties are shown as the uncertainties related to the bias correction can not be propagated from the BDT classifier
to a different variable.}
\label{fig:btagCR}
\end{figure}

\begin{figure}[h!tbp]
\includegraphics[width=1\textwidth]{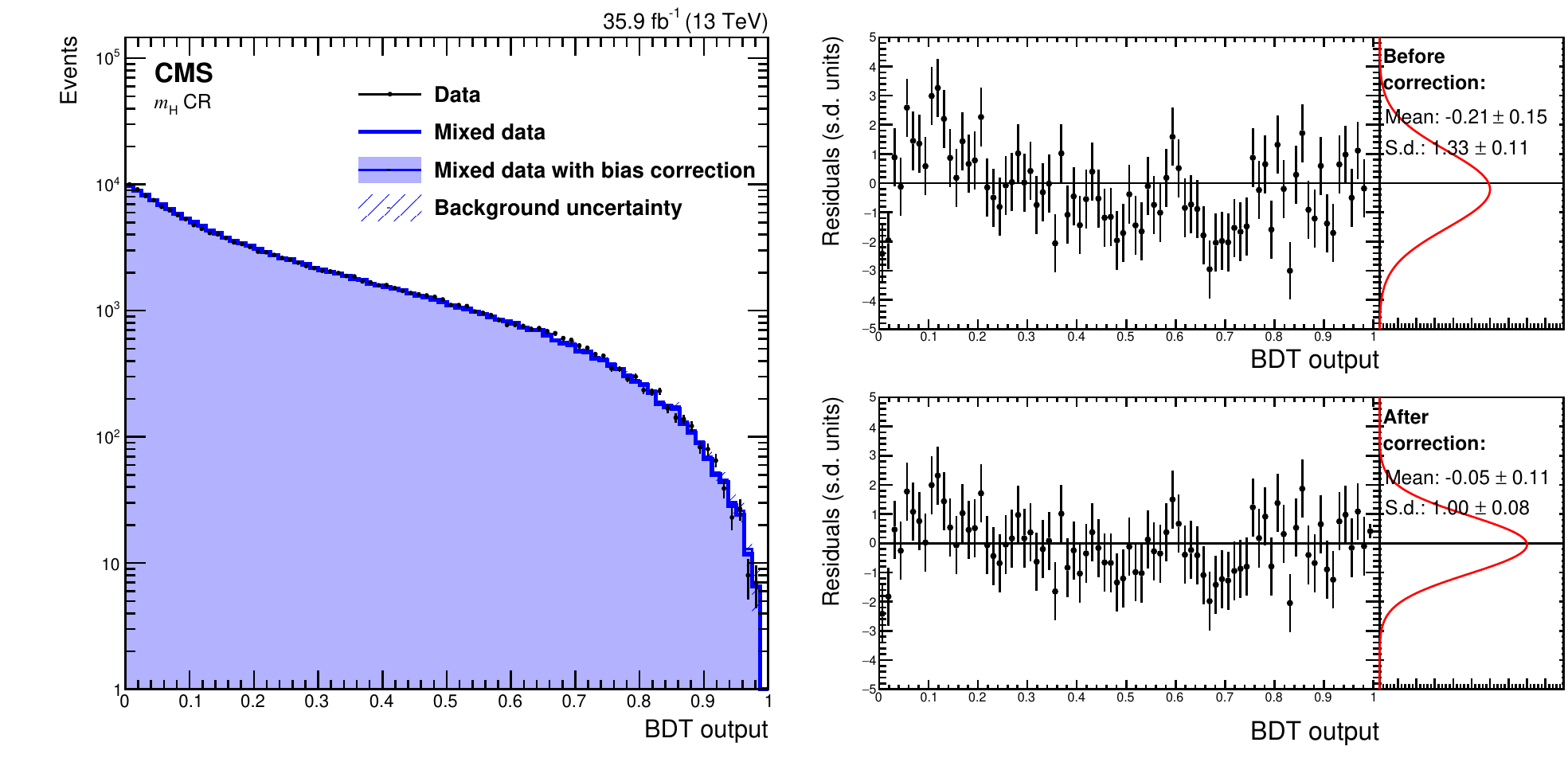} \\
\caption {Left: comparison of the distribution of BDT output for data (left)
selected in a region of the leading versus trailing Higgs boson candidate mass plane that excludes
a 60-\GeV-wide box around the most probable values of the dijet masses of signal events,
with the corresponding output on an artificial sample obtained from the same data set by hemisphere mixing.
Right: bin-by-bin differences between data and model,
in \sd units before (upper right) and after (lower right) bias correction; pull distribution for the differences, fit to a Gaussian distribution.
The bias correction uncertainty is increased to take the \sd of the residuals to 1.0.}
\label{fig:check1}
\end{figure}

We perform a number of stringent checks to verify that the background is well modelled by the hemisphere mixing procedure.
For this purpose, we define two control regions (CRs):
the first one, called the \mHiggs\ CR, is obtained by removing from the data events where the leading and trailing dijet masses
are in the region $90 < \MHone < 150\GeV$, $80 < \MHtwo < 140\GeV$.
This avoids using events belonging to the signal-enriched region.
In the second region, the \cPqb\ tag CR, fully orthogonal to the default selection,
we select events with at least four \cPqb-tagged jets as defined by the loose working point of the cMVAv2,
while vetoing events with any jets that are defined as \cPqb-tagged jets according to the medium working point of the cMVAv2.
The loose working point of the cMVAv2 has a misidentification rate of ${\approx}10\%$ and
a \cPqb-tagging efficiency of ${\approx}85\%$ for jets
produced by the hadronization of \cPqb\ quarks emitted in \HH production events.
The distributions of all individual event variables for the artificial data sets are compared to those from the original data set
in these two CRs and are found to be in agreement.
This is illustrated for a number of variables in Figs.~\ref{fig:massCR} and~\ref{fig:btagCR}.
However, the power of the technique rests in its ability to provide fully multidimensional modelling.
To verify this, a first cross-check consists of comparing the full BDT shape for data and the artificial model in the \mHiggs\ CR.
We observe an agreement in the shape of the BDT discriminator with a slight excess of background events in the lower range of the BDT output
(as can be seen in Fig.~\ref{fig:check1}, left). A similar trend is seen in the \cPqb\ tag CR.


A high-precision study is required to investigate the need for a correction to the background shape of the BDT discriminator and a corresponding systematic uncertainty.
For this purpose, all the possible combinations of neighbouring hemispheres in the range $1$ to $10$,
except the ones used for training ($(1,1),(1,2),(2,1),(2,2)$),
are merged into a unique sample $M$.
We re-sample 200 new replicas with the same number of events as the original data set without replacement from $M$,
each time starting from the full sample $M$.
Each of the replicas is then used as a new original data set, and artificial samples are created from it using the hemisphere mixing procedure.
The output distribution of a previously trained BDT for the large sample $M$ is then compared to that for its artificial counterpart,
obtaining a distribution of differences between actual and predicted data in each of the 80 BDT bins.
A schematic of the procedure and the results are available in~\suppMaterial.
A systematic bias is detected and the background template is corrected for the value obtained from this comparison.
The variance related to the background bias extraction, together with expected statistical uncertainty,
are estimated and accounted for as a systematic uncertainty in the final fit described in Section~\ref{sec:res}.
The validity of this background bias extraction procedure has been checked by applying it to the data in the two CRs previously mentioned.
The means of the per bin expectation values minus the observed values are compatible with zero after the bias correction in both control regions,
the root-mean-square of the pulls is compatible with one after the bias correction in the \cPqb\ tag CR,
but not in the \mHiggs\ CR, as shown on Fig.~\ref{fig:check1} (upper right).
To account for this, we increase the uncertainty in the background such that the value of standard deviation (s.d.) becomes 1.0 in the \mHiggs\ CR after the bias correction is applied (Fig.~\ref{fig:check1}, lower right).

\section{Systematic uncertainties}
\label{sec:systematics}
The sources of systematic uncertainties found to be relevant to this analysis
are listed in Table~\ref{tab:sys}.
The systematic uncertainty in the shape of the background model is accounted for by assigning an uncertainty to each BDT output bin that includes the statistical uncertainty and the systematic uncertainty related to the bias extraction discussed in the previous section.
The background normalization is left freely floating in the BDT distribution fit.
The uncertainty due to the \cPqb\ tagging efficiencies is estimated by varying them within their uncertainties.
The uncertainty due to the pileup modelling is computed by considering a ${\pm}4.6\%$ variation
in the total inelastic cross section value at 13\TeV~\cite{Sirunyan:2018nqx}.
The effect of jet energy resolution is evaluated by smearing
jet energies according to their estimated uncertainty.
The jet energy scale is varied within one \sd\ as a function of jet \pt and $\abs{\eta}$,
and the efficiency of the selection criteria is recomputed.
The trigger efficiency correction factor discussed in Section~\ref{s:event_reco} is
affected by a 2\% uncertainty
that is taken as a systematic uncertainty in the related source.
In the mentioned sources of systematic uncertainty, both shape and normalization shifts are considered in the model.
The signal yield for a given production cross section is affected by a systematic uncertainty in the measured
integrated luminosity of 2.5\%~\cite{CMS-PAS-LUM-17-001}.
The effect of variation of the \muR\ and \muF\ scales
on the signal acceptance is estimated by taking the maximum and the
minimum difference with respect to the nominal acceptance when varying
\muF\ and \muR\ each individually as well as both together up and down by a factor of two.
Lastly, to estimate the signal acceptance uncertainty due to PDF uncertainties, the PDF4LHC~\cite{Butterworth:2015oua}
recommendation is followed, using as the uncertainty the \sd in the acceptance for a set of 100 MC replicas
of the NNPDF~3.0 set~\cite{nnpdf}.

\begin{table}[htbp]
 \topcaption{Systematic uncertainties considered in the analysis
   and relative impact on the expected limit for the SM \HH production. The relative impact is obtained
   by fixing the nuisance parameters corresponding to each source and recalculating the expected limit.}
 \centering
 \begin{tabular}{l c c}
   \hline
   Source        & Affects  & Exp. limit variation \\
   \hline
   Bkg. shape     & bkg.  & 30\%  \\
   Bkg. norm.     & bkg.  & 8.6\%  \\[\cmsTabSkip]
   \cPqb\ tagging eff.& sig  & 2.8\%  \\
   Pileup        & sig  &       ${<} 0.01\%$ \\
   Jet energy res.  & sig  &    ${<} 0.01\%$ \\
   Jet energy scale  & sig  &  ${<} 0.01\%$ \\
   Int. luminosity    & sig  & ${<} 0.01\%$ \\
   Trigger eff.  & sig  &      ${<} 0.01\%$ \\
   \muF\ and \muR\ scales  & sig  & ${<} 0.01\%$ \\
   PDF           & sig  &       ${<} 0.01\%$ \\
   \hline
 \end{tabular}
 \label{tab:sys}
\end{table}

\section{Results}
\label{sec:res}
\begin{figure}[htbp]
\centering
\includegraphics[width=0.75\textwidth]{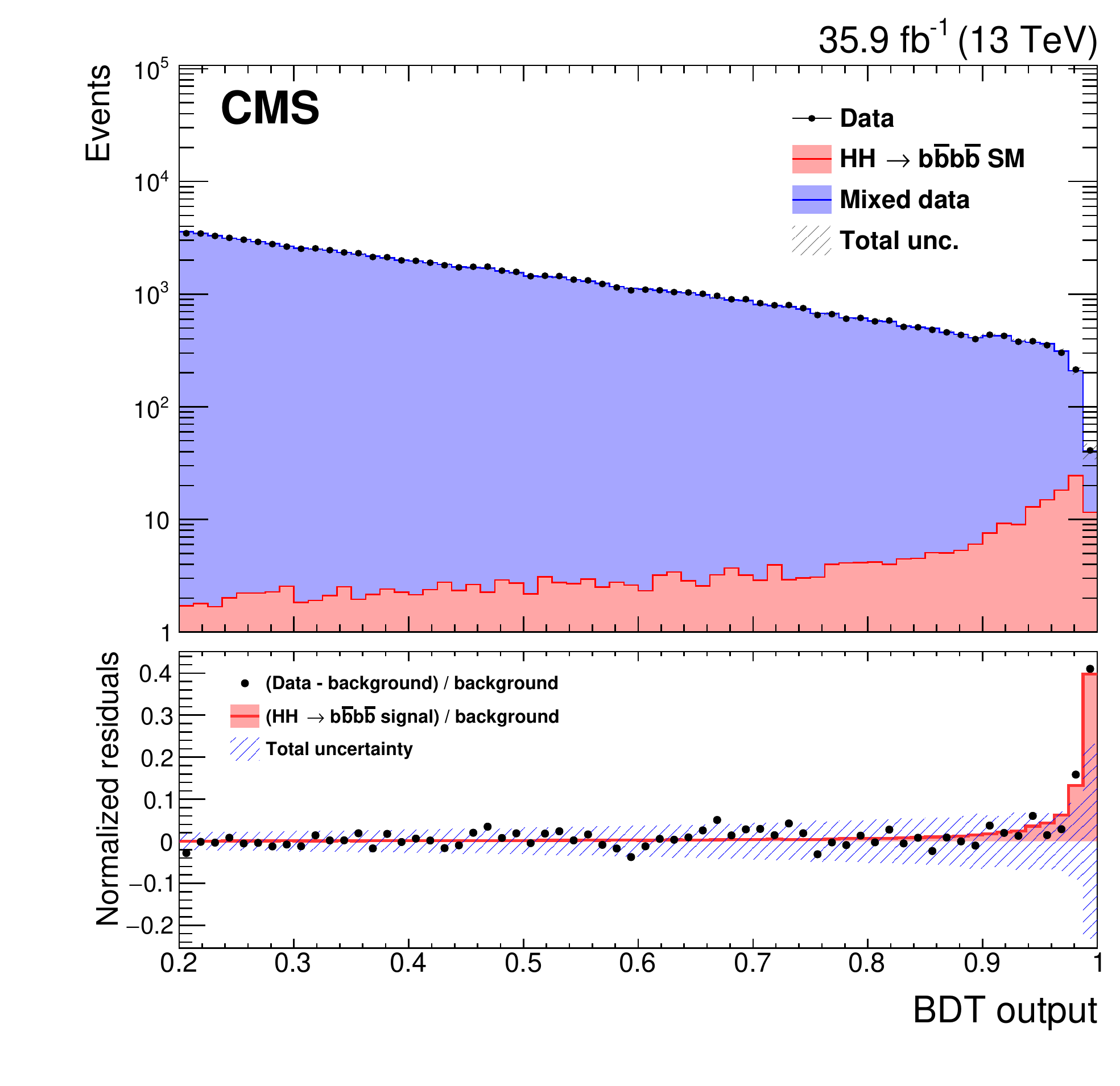}
\caption{Results of the fit to the BDT distribution for the SM \HH production signal. In the bottom panel a comparison is shown between
the best fit signal and best fit background subtracted from measured data. The band, centred at zero, shows the total uncertainty.}
\label{fig:SMfit}
\end{figure}

\begin{figure}[htbp]
\centering
\includegraphics[width=0.49\textwidth]{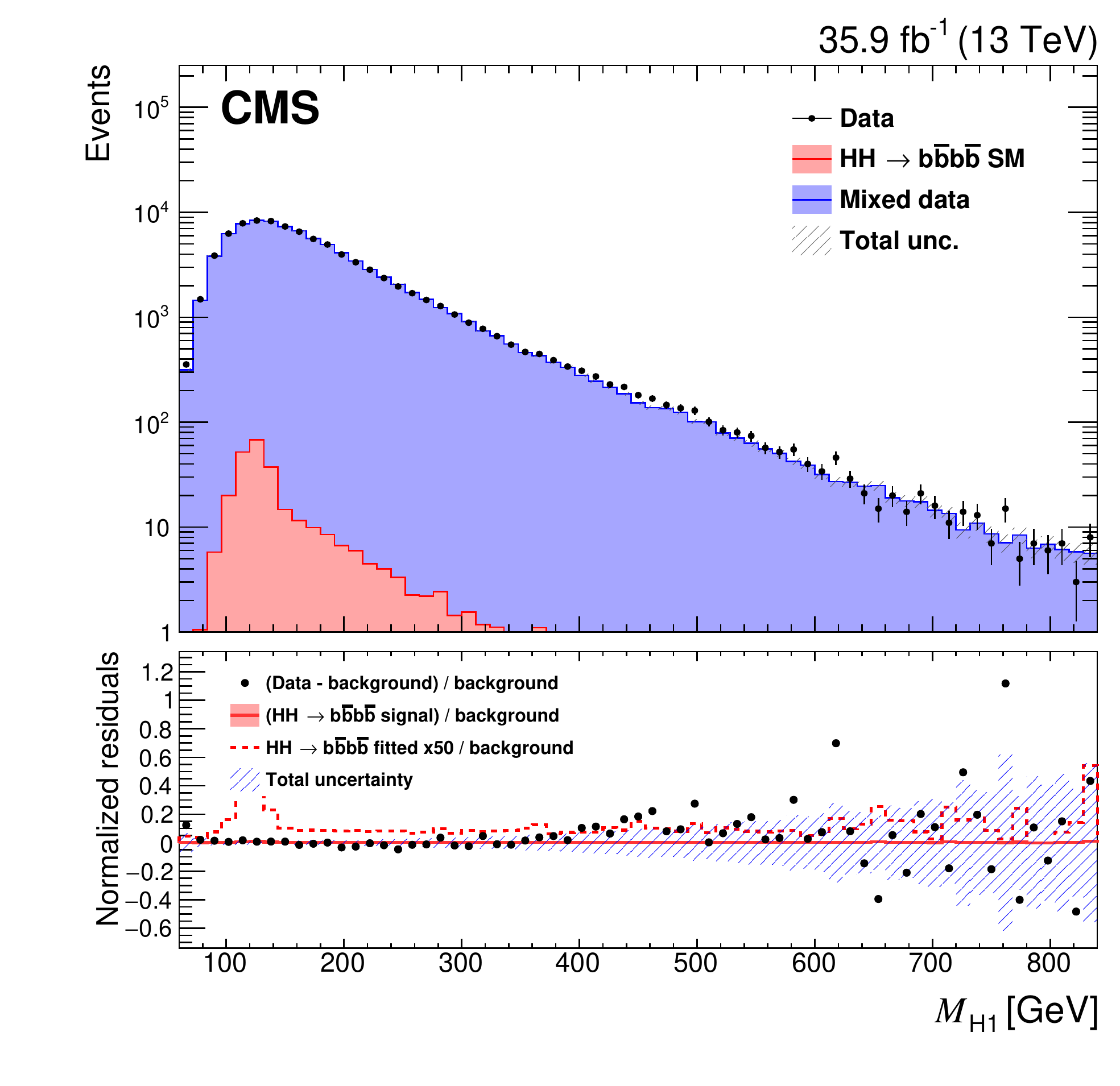}
\includegraphics[width=0.49\textwidth]{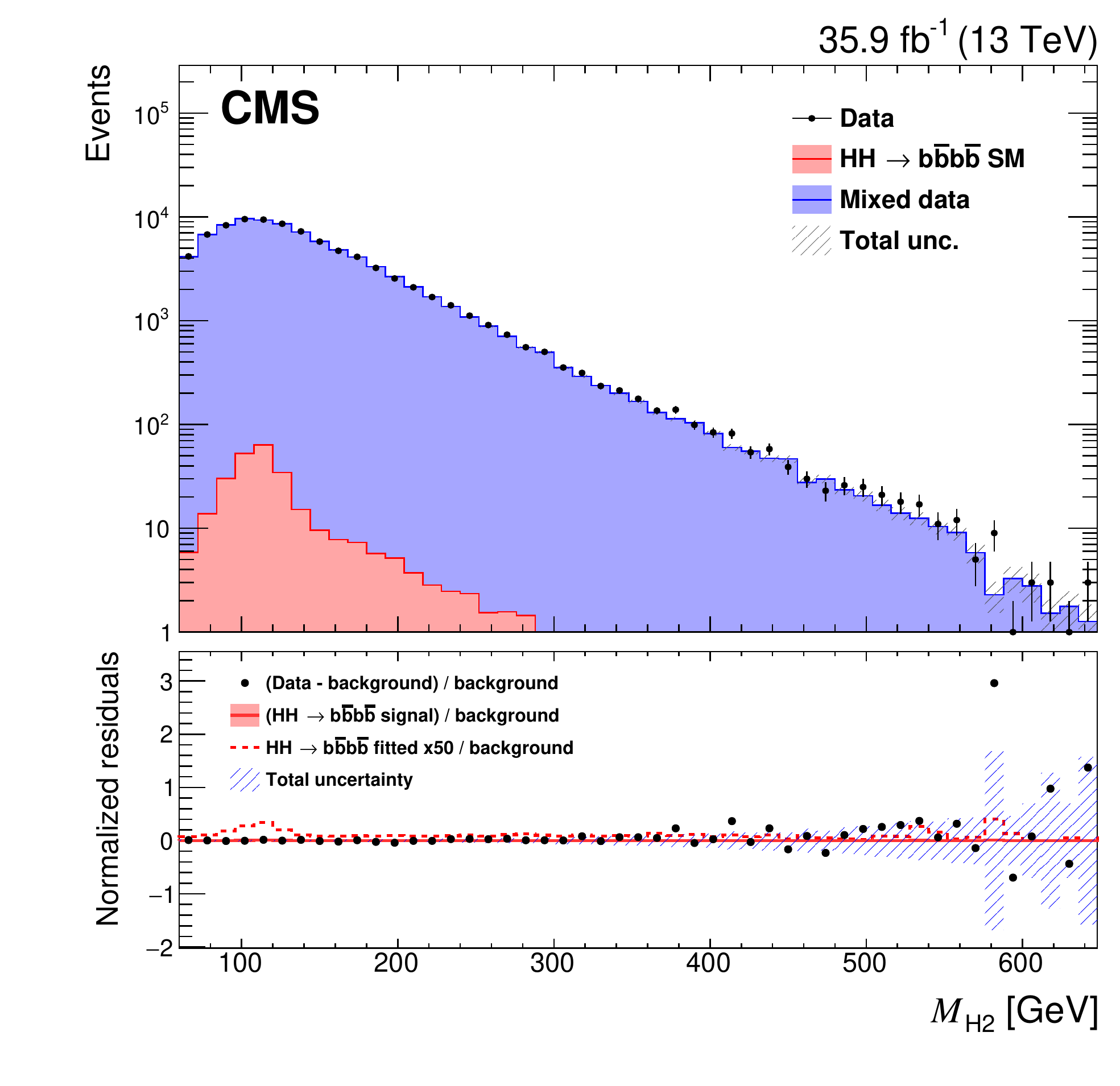}  \\
\caption{Post-fit distribution of \MHone\ (left) and \MHtwo\ (right).
Bias correction for the background model is applied by rescaling the weight of each event using the event yield ratio between corrected and uncorrected BDT distributions.}
\label{fig:postfit}
\end{figure}

We search for the presence of \HH events in CMS data collected in the 2016
run of the LHC using the BDT discriminant trained on the SM signal simulation and artificial background data.
Two-component likelihood fits to the binned BDT output distributions are performed,
using the BDT distribution for the background resulting from the artificial data set described in Section~\ref{sec:background}
and the signal simulations corresponding to the SM and each of the BSM benchmark points.
The validation samples were used to study the dependence of both the expected limit and the compatibility of the data and background distributions on the value of the BDT discriminator used for the selection.
Selecting  BDT discriminator values ${>} 0.2$ results in a small loss of sensitivity (${\approx}1.5\%$) with improved data-background compatibility.
As a result, the 64 bins with BDT ${>} 0.2$ are used to extract the limits.
The fit to the SM signal is shown in Fig.~\ref{fig:SMfit} and the postfit distributions of reconstructed Higgs boson masses are shown in Fig.~\ref{fig:postfit}.
Minor background contamination arising from
\ttH, \ZH, \bbH,
and single Higgs boson production processes do not show a signal-like BDT distribution and their effect is found to be negligible in the selected data at our level of sensitivity.

The observed and expected 95\% confidence level (\CL) upper limits on the cross section for
\ppToHHbbbb nonresonant production, are computed using the asymptotic approximation~\cite{Cowan:2010js} of the \CLs criterion~\cite{Read:2002hq, CLS2, CMS-NOTE-2011-005}, using a test statistic based on the profile likelihood ratio (the LHC test statistic)~\cite{Cowan:2010js}.
The systematic uncertainties are treated as nuisance parameters and are profiled in the minimization.
The limits are shown in Table~\ref{tab:limits2} together with the 1\sd and 2\sd \CL intervals around the expected limits.
For the SM process, the expected limit is 419\fb, which corresponds to ${\approx}37$ times the SM \HH production cross section times
the square of the branching fraction for the \HTobb decay.
The observed upper limit obtained is 847\fb, which is ${\approx}2\sd$ above the expected upper limit.
This corresponds to an observed limit of 2496\fb for \xsppToHHSM.

\begin{table}[htbp]
 \topcaption{The observed and expected upper limits on \xsppToHHbbbb in the SM
at 95\% \CL in units of fb.}
 \centering
 \begin{tabular}{l l l l l l l}
   \hline
   Category      & Observed & Expected & -2\sd & -1\sd & +1\sd & +2\sd \\
   \hline
SM \HHTobbbb &   847 &   419  &   221  &   297  &   601  &   834  \\
   \hline
 \end{tabular}
 \label{tab:limits2}
\end{table}

\begin{figure}[hp]
\centering
\includegraphics[width=0.65\textwidth]{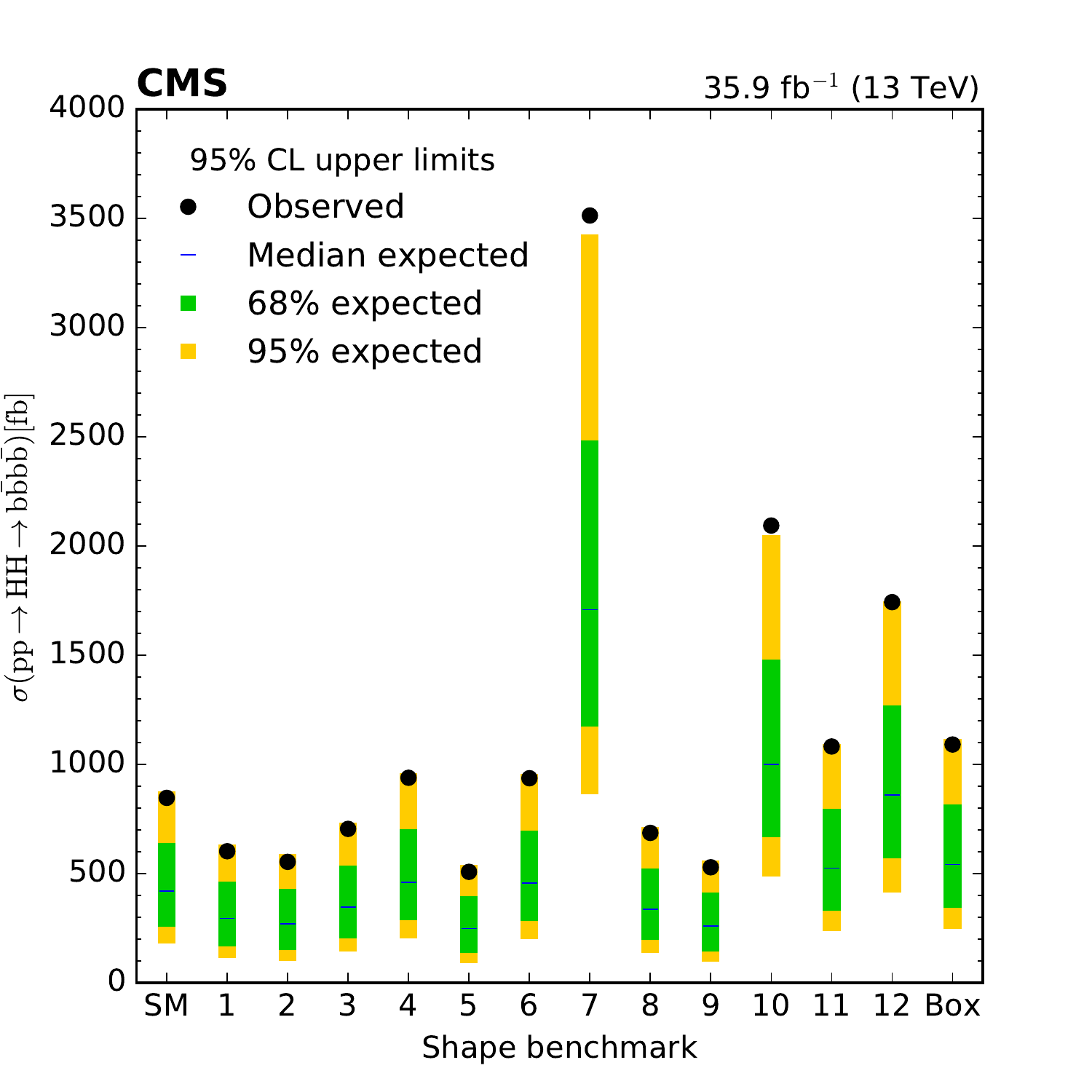}
\caption {The observed and expected upper limits at 95\% \CL on the \xsppToHHbbbb cross section for the 13 BSM models investigated.
See Table~\ref{table:benchmarks} for their respective parameter values.}
\label{f:bsmlimits}
\end{figure}

\begin{figure}[hp]
\centering
\includegraphics[width=0.65\textwidth]{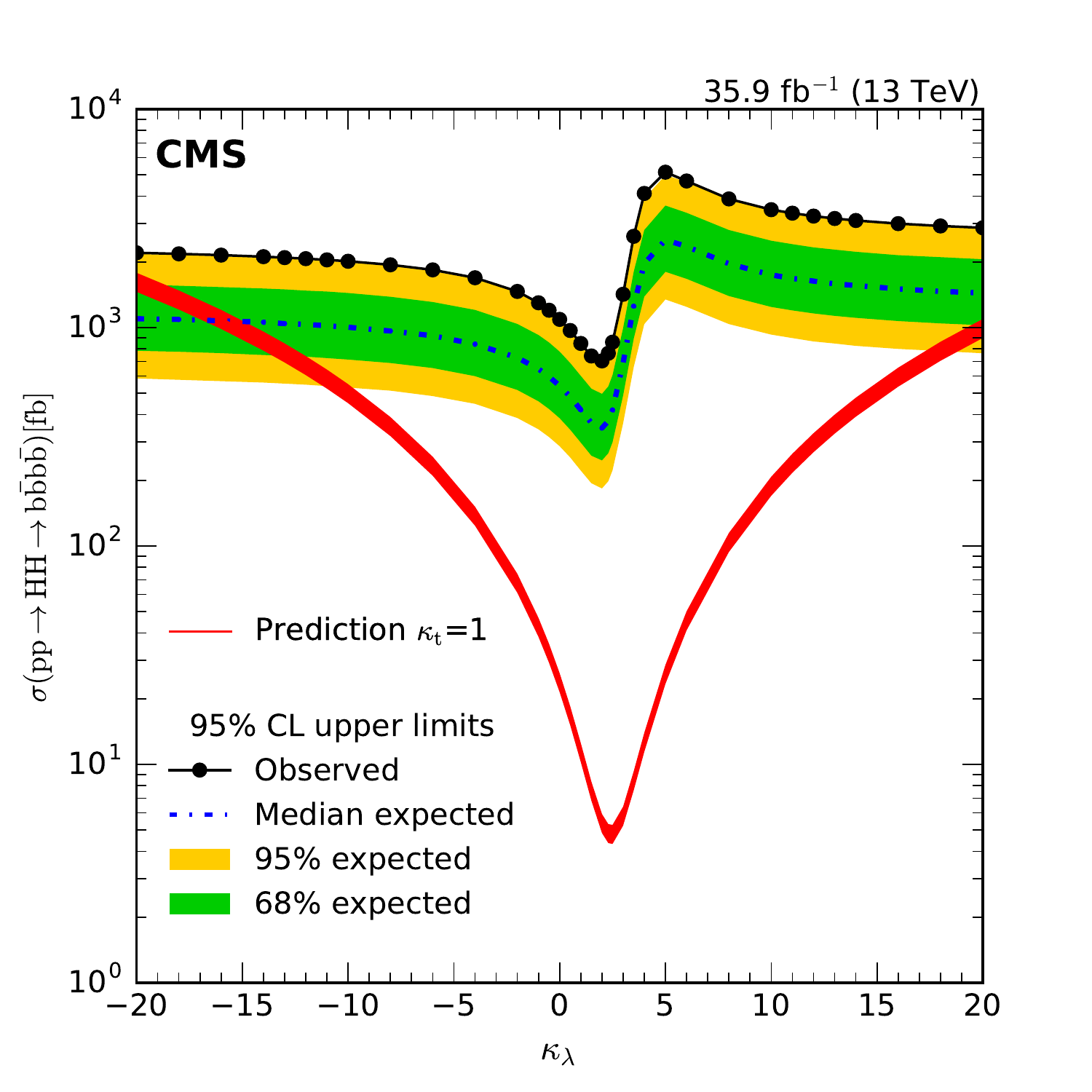}
\caption {95\% \CL cross section limits on \xsppToHHbbbb for values of \kappalambda in the [-20,20] range, assuming  $\kappat = 1$;
the theoretical prediction with $\kappat = 1$ is also shown.}
\label{fig:lim_bsm_scan}
\end{figure}

We perform the procedure described above in turn on the 13 BSM benchmark models considered.
The results are shown in Fig.~\ref{f:bsmlimits} and reported in Table~\ref{t:limits}.
The difference between observed and expected limits is similar for SM and all the benchmark models.
This is explained by the fact that the benchmark points use the same BDT as SM, resulting in the same background shape as an input to the fit.
The background shape has a deficit of events compared to data in the last bins of the BDT distribution, as seen in Fig.~\ref{fig:SMfit}.
We also search for \HH production with values of \kappalambda in the range [-20, 20], assuming  $\kappat = 1$, and the results are shown in Fig.~\ref{fig:lim_bsm_scan}.
The kinematic properties vary significantly across the points in this range.
We do not exclude any values of \kappalambda, assuming $\kappat = 1$.

\begin{table}[htbp]
 \centering
\topcaption{The observed and expected upper limits on the \xsppToHHbbbb cross section for the 13 BSM benchmark models at 95\% \CL in units of fb.}
 \begin{tabular}{l l l l l l l }
\hline
Benchmark point & Observed & Expected & -2\sd & -1\sd & +1\sd & +2\sd  \\
\hline
1 &   602  &   295  &   155  &   209  &   424  &   592  \\
2 &   554  &   269  &   141  &   190  &   389  &   548  \\
3 &   705  &   346  &   182  &   245  &   497  &   691  \\
4 &   939  &   461  &   244  &   327  &   662  &   920  \\
5 &   508  &   248  &   131  &   176  &   357  &   501  \\
6 &   937  &   457  &   240  &   323  &   657  &   916  \\
7 &   3510  &   1710  &   905  &   1210  &   2440  &   3390  \\
8 &   686  &   336  &   177  &   238  &   483  &   674  \\
9 &   529  &   259  &   136  &   183  &   373  &   520  \\
10 &   2090  &   1000  &   527  &   709  &   1440  &   2010  \\
11 &   1080  &   525  &   277  &   372  &   755  &   1050  \\
12 &   1744  &   859  &   455  &   611  &   1230  &   1710  \\
Box &   1090  &   542  &   286  &   384  &   775  &   1080  \\
   \hline
\end{tabular}
\label{t:limits}
\end{table}

\section{Summary}
This paper presents a search for nonresonant Higgs boson pair (\HH) production with both Higgs bosons decaying into \bbbar pairs.
The standard model (SM) production has been studied along with 13 beyond the SM (BSM) benchmark models,
using a data set of $\sqrt{s} = 13\TeV$ proton-proton collision events,
corresponding to an integrated luminosity of 35.9\fbinv collected by the CMS detector during the 2016 LHC run.
The analysis of events acquired by a hadronic multijet trigger includes the selection of events with 4 \cPqb-tagged jets
and a classification using boosted decision trees, optimized for discovery of the SM \HH signal.
Limits at 95\% confidence level on the \HH production cross section times the square of the branching fraction for the Higgs boson decay to \cPqb\ quark pairs
are extracted for the SM and each BSM model considered, using binned likelihood fits of the shape of the boosted decision tree classifier output.
The background model is derived
from a novel technique based on data that provides a multidimensional representation of
 the dominant quantum chromodynamics multijet background and also models well the overall background distribution.
The expected upper limit on \xsppToHHbbbb is 419\fb, corresponding to 37 times the expected value for the SM process.
The observed upper limit is 847\fb.
Anomalous couplings of the Higgs boson are also investigated.
The upper limits extracted for the \HH production cross section in the 13 BSM benchmark models range from
508 to 3513\fb.

\begin{acknowledgments}
We congratulate our colleagues in the CERN accelerator departments for the excellent performance of the LHC and thank the technical and administrative staffs at CERN and at other CMS institutes for their contributions to the success of the CMS effort. In addition, we gratefully acknowledge the computing centres and personnel of the Worldwide LHC Computing Grid for delivering so effectively the computing infrastructure essential to our analyses. Finally, we acknowledge the enduring support for the construction and operation of the LHC and the CMS detector provided by the following funding agencies: BMBWF and FWF (Austria); FNRS and FWO (Belgium); CNPq, CAPES, FAPERJ, FAPERGS, and FAPESP (Brazil); MES (Bulgaria); CERN; CAS, MoST, and NSFC (China); COLCIENCIAS (Colombia); MSES and CSF (Croatia); RPF (Cyprus); SENESCYT (Ecuador); MoER, ERC IUT, and ERDF (Estonia); Academy of Finland, MEC, and HIP (Finland); CEA and CNRS/IN2P3 (France); BMBF, DFG, and HGF (Germany); GSRT (Greece); NKFIA (Hungary); DAE and DST (India); IPM (Iran); SFI (Ireland); INFN (Italy); MSIP and NRF (Republic of Korea); MES (Latvia); LAS (Lithuania); MOE and UM (Malaysia); BUAP, CINVESTAV, CONACYT, LNS, SEP, and UASLP-FAI (Mexico); MOS (Montenegro); MBIE (New Zealand); PAEC (Pakistan); MSHE and NSC (Poland); FCT (Portugal); JINR (Dubna); MON, RosAtom, RAS, RFBR, and NRC KI (Russia); MESTD (Serbia); SEIDI, CPAN, PCTI, and FEDER (Spain); MOSTR (Sri Lanka); Swiss Funding Agencies (Switzerland); MST (Taipei); ThEPCenter, IPST, STAR, and NSTDA (Thailand); TUBITAK and TAEK (Turkey); NASU and SFFR (Ukraine); STFC (United Kingdom); DOE and NSF (USA).

\hyphenation{Rachada-pisek} Individuals have received support from the Marie-Curie programme and the European Research Council and Horizon 2020 Grant, contract No. 675440 (European Union); the Leventis Foundation; the A. P. Sloan Foundation; the Alexander von Humboldt Foundation; the Belgian Federal Science Policy Office; the Fonds pour la Formation \`a la Recherche dans l'Industrie et dans l'Agriculture (FRIA-Belgium); the Agentschap voor Innovatie door Wetenschap en Technologie (IWT-Belgium); the F.R.S.-FNRS and FWO (Belgium) under the ``Excellence of Science - EOS" - be.h project n. 30820817; the Ministry of Education, Youth and Sports (MEYS) of the Czech Republic; the Lend\"ulet (``Momentum") Programme and the J\'anos Bolyai Research Scholarship of the Hungarian Academy of Sciences, the New National Excellence Program \'UNKP, the NKFIA research grants 123842, 123959, 124845, 124850 and 125105 (Hungary); the Council of Science and Industrial Research, India; the HOMING PLUS programme of the Foundation for Polish Science, cofinanced from European Union, Regional Development Fund, the Mobility Plus programme of the Ministry of Science and Higher Education, the National Science Center (Poland), contracts Harmonia 2014/14/M/ST2/00428, Opus 2014/13/B/ST2/02543, 2014/15/B/ST2/03998, and 2015/19/B/ST2/02861, Sonata-bis 2012/07/E/ST2/01406; the National Priorities Research Program by Qatar National Research Fund; the Programa Estatal de Fomento de la Investigaci{\'o}n Cient{\'i}fica y T{\'e}cnica de Excelencia Mar\'{\i}a de Maeztu, grant MDM-2015-0509 and the Programa Severo Ochoa del Principado de Asturias; the Thalis and Aristeia programmes cofinanced by EU-ESF and the Greek NSRF; the Rachadapisek Sompot Fund for Postdoctoral Fellowship, Chulalongkorn University and the Chulalongkorn Academic into Its 2nd Century Project Advancement Project (Thailand); the Welch Foundation, contract C-1845; and the Weston Havens Foundation (USA).
\end{acknowledgments}

\bibliography{auto_generated}
\appendix
\section{Supplemental material\label{app:suppMat}}
\input{supplemental}

\cleardoublepage \section{The CMS Collaboration \label{app:collab}}\begin{sloppypar}\hyphenpenalty=5000\widowpenalty=500\clubpenalty=5000\input{HIG-17-017-authorlist.tex}\end{sloppypar}
\end{document}

%% file: supplemental.tex
\begin{figure}[h!]
\centering
\includegraphics[width=0.8\textwidth]{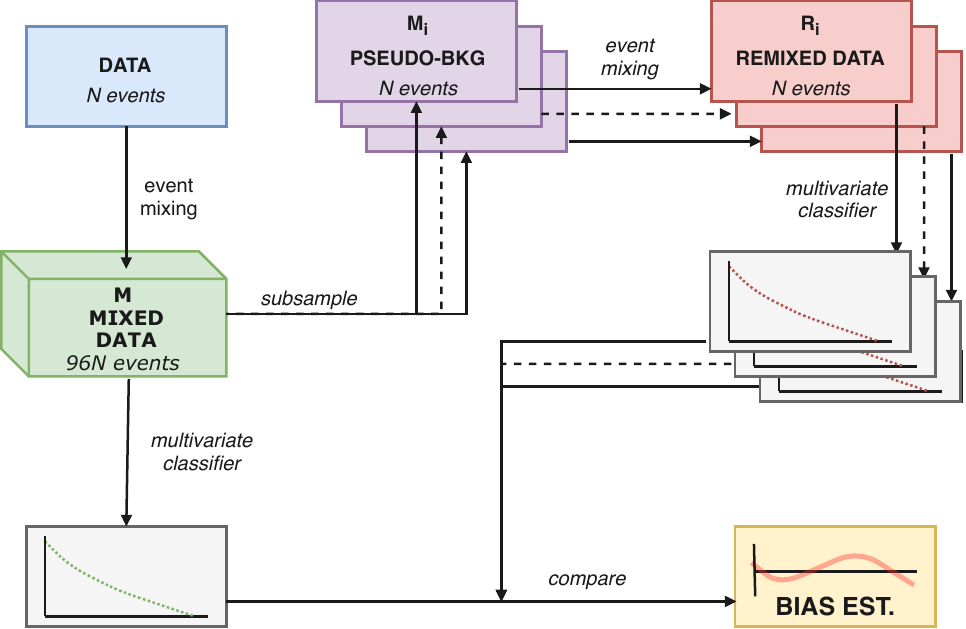}
\caption {Diagram describing the procedure used to estimate the
background bias correction. All possible combinations of
mixed hemispheres except those used for training are added together to
create a large sample $M$ of $96N$ events
from which we repeatedly subsample without
replacement 200 replicas $M_i$ of $N$ events.
The hemisphere mixing procedure is then
carried out again for each of this replicas to produce a set of
re-mixed data replicas $R_i$. The trained multivariate classifier
trained is then evaluated over all the events of $M$ and each $R_i$.
and the histograms of the classifier output are compared to
obtain a the differences for each of the replicas. The median
difference is taken as bias correction.}
\label{f:bias_extraction}
\end{figure}

\begin{figure}[h!]
\centering
\includegraphics[width=1.0\textwidth]{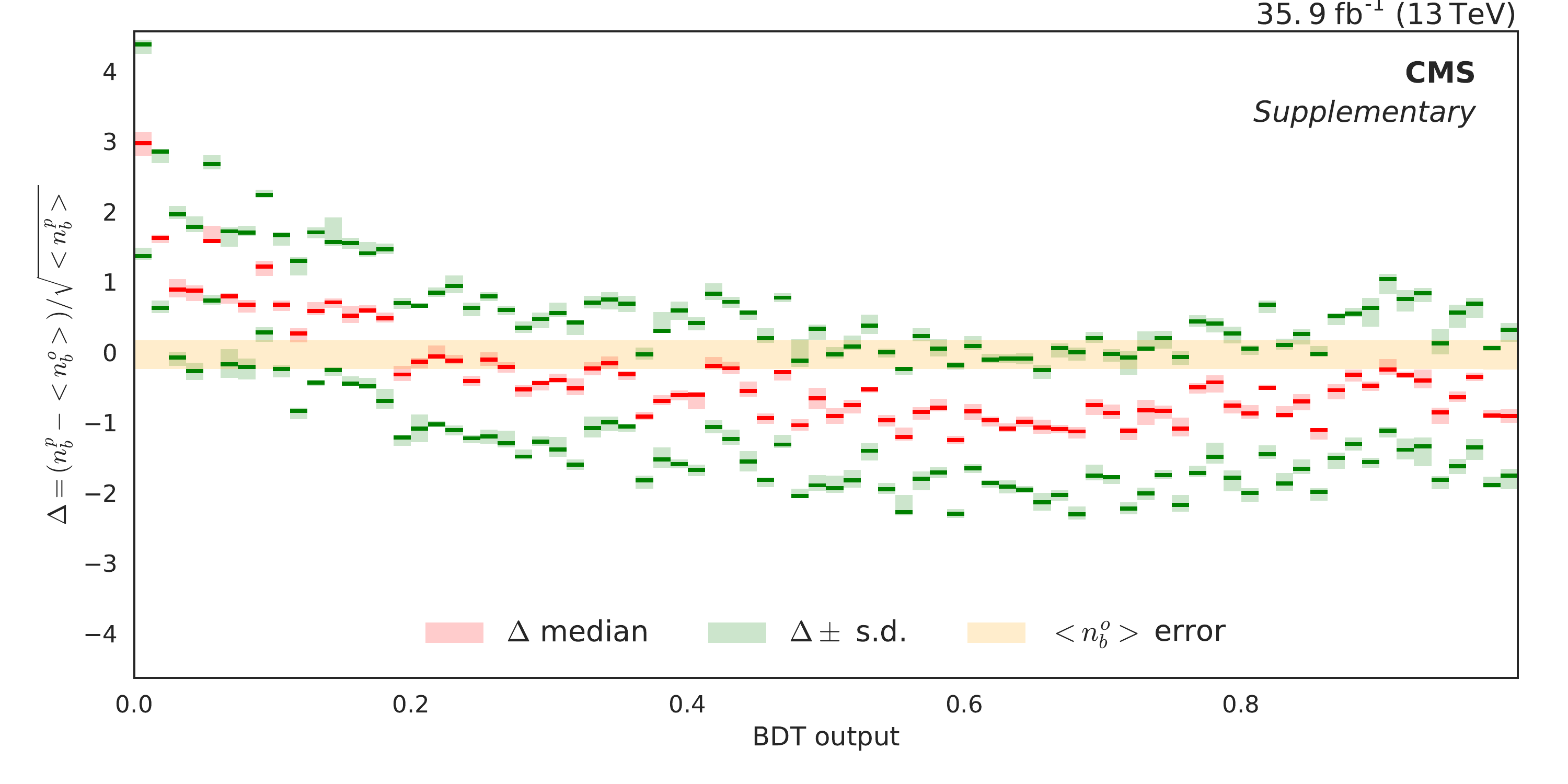}
\caption {Bias estimation by resampling, in relative units of the statistical
uncertainty of the predicted background, used to
correct the background estimation. The median (red line)
and the upper and lower one \sd quantiles (green lines) have been computed
from 200 subsamples of the re-mixed data comparing the predicted background
$n^p_b$ with the observed $n^o_b$. The variability due to the limited number of
subsamples is estimated by bootstrap and it is shown for each estimation using
a coloured shadow around the quantile estimation.
The light yellow shadow represents the uncertainty
due to the limited statistics of the reference observed sample.
The separation between the one \sd quantiles is compatible with the
expected variance if the estimation was Poisson or Gaussian distributed.
}
\label{f:check3}
\end{figure}

%% file: HIG-17-017-authorlist.tex
\vskip\cmsinstskip
\textbf{Yerevan Physics Institute, Yerevan, Armenia}\\*[0pt]
A.M.~Sirunyan, A.~Tumasyan
\vskip\cmsinstskip
\textbf{Institut f\"{u}r Hochenergiephysik, Wien, Austria}\\*[0pt]
W.~Adam, F.~Ambrogi, E.~Asilar, T.~Bergauer, J.~Brandstetter, M.~Dragicevic, J.~Er\"{o}, A.~Escalante~Del~Valle, M.~Flechl, R.~Fr\"{u}hwirth\cmsAuthorMark{1}, V.M.~Ghete, J.~Hrubec, M.~Jeitler\cmsAuthorMark{1}, N.~Krammer, I.~Kr\"{a}tschmer, D.~Liko, T.~Madlener, I.~Mikulec, N.~Rad, H.~Rohringer, J.~Schieck\cmsAuthorMark{1}, R.~Sch\"{o}fbeck, M.~Spanring, D.~Spitzbart, A.~Taurok, W.~Waltenberger, J.~Wittmann, C.-E.~Wulz\cmsAuthorMark{1}, M.~Zarucki
% ...
\vskip\cmsinstskip
% ...
1:  Also at Vienna University of Technology, Vienna, Austria\\
% ...
